\documentclass[twocolumn,prc,aps,showpacs,superscriptaddress,amsmath,amssymb,floatfix,nofootinbib,superscriptaddress]{revtex4}
\usepackage[usenames]{color}
\usepackage{graphicx}
\usepackage{dcolumn}
\usepackage{bm}

\begin{document}


\title{Study of the A(e,e'$\pi^+$) Reaction on $^1$H, $^2$H, $^{12}$C, $^{27}$Al,
  $^{63}$Cu and $^{197}$Au}

\author{X.~Qian}
\affiliation{Triangle Universities Nuclear Laboratory, Duke University, Durham, 
  NC, USA}
\author{T.~Horn}
\affiliation{University of Maryland, College Park, MD, USA}
\affiliation{Physics Division, TJNAF, Newport News, VA, USA}
\author{B.~Clasie}
\affiliation{Laboratory for Nuclear Science, Massachusetts Institute of Technology, 
  Cambridge, MA, USA}
\author{J.~Arrington}
\affiliation{Argonne National Laboratory, Argonne, IL, USA}
\author{R.~Asaturyan}
\affiliation{Yerevan Physics Institute, Yerevan, Armenia}
\author{F.~Benmokhtar}
\affiliation{University of Maryland, College Park, MD, USA}
\author{W.~Boeglin}
\affiliation{Florida International University, Miami, FL, USA}
\author{P.~Bosted}
\affiliation{Physics Division, TJNAF, Newport News, VA, USA}
\author{A.~Bruell}
\affiliation{Physics Division, TJNAF, Newport News, VA, USA}
\author{M.~E.~Christy}
\affiliation{Hampton University, Hampton, VA, USA}
\author{E.~Chudakov}
\affiliation{Physics Division, TJNAF, Newport News, VA, USA}
\author{M.~M.~Dalton}
\affiliation{University of the Witwatersrand, Johannesburg, South Africa}
\author{A.~Daniel}
\affiliation{University of Houston, Houston, TX, USA}
\author{D.~Day}
\affiliation{University of Virginia, Charlottesville, VA, USA}
\author{D.~Dutta}
\affiliation{Mississippi State University, Mississippi State, MS, USA}
\affiliation{Triangle Universities Nuclear Laboratory, Duke University, Durham, 
  NC, USA}
\author{L.~El~Fassi}
\affiliation{Argonne National Laboratory, Argonne, IL, USA}
\author{R.~Ent}
\affiliation{Physics Division, TJNAF, Newport News, VA, USA}
\author{H.~C.~Fenker}
\affiliation{Physics Division, TJNAF, Newport News, VA, USA}
\author{J.~Ferrer}
\affiliation{James Madison University, Harrisonburg, VA, USA}
\author{N.~Fomin}
\affiliation{University of Virginia, Charlottesville, VA, USA}
\author{H.~Gao}
\affiliation{Triangle Universities Nuclear Laboratory, Duke University, Durham, 
  NC, USA}
\author{K.~Garrow}
\affiliation{TRIUMF, Vancouver, British Columbia, Canada}
\author{D.~Gaskell}
\affiliation{Physics Division, TJNAF, Newport News, VA, USA}
\author{C.~Gray}
\affiliation{University of the Witwatersrand, Johannesburg, South Africa}
\author{G.~M.~Huber}
\affiliation{University of Regina, Regina, Saskatchewan, Canada}
\author{M.~K.~Jones}
\affiliation{Physics Division, TJNAF, Newport News, VA, USA}
\author{N.~Kalantarians}
\affiliation{University of Houston, Houston, TX, USA}
\author{C.~E.~Keppel}
\affiliation{Hampton University, Hampton, VA, USA}
\affiliation{Physics Division, TJNAF, Newport News, VA, USA}
\author{K.~Kramer}
\affiliation{Triangle Universities Nuclear Laboratory, Duke University, Durham, 
  NC, USA}
\author{Y.~Li}
\affiliation{University of Houston, Houston, TX, USA}
\author{Y.~Liang}  
\affiliation{Ohio University, Athens, OH, USA}
\author{A.~F.~Lung}
\affiliation{Physics Division, TJNAF, Newport News, VA, USA}
\author{S.~Malace}
\affiliation{Hampton University, Hampton, VA, USA}
\author{P.~Markowitz}
\affiliation{Florida International University, Miami, FL, USA}
\author{A.~Matsumura}
\affiliation{Tohoku University, Sendai, Japan}
\author{D.~G.~Meekins}
\affiliation{Physics Division, TJNAF, Newport News, VA, USA}
\author{T.~Mertens}
\affiliation{Basel University, Basel, Switzerland}
\author{T.~Miyoshi}
\affiliation{University of Houston, Houston, TX, USA}
\author{H.~Mkrtchyan}
\affiliation{Yerevan Physics Institute, Armenia}
\author{R.~Monson}
\affiliation{Central Michigan University, Mount Pleasant, MI, USA}
\author{T.~Navasardyan}
\affiliation{Yerevan Physics Institute, Armenia}
\author{G.~Niculescu}
\affiliation{James Madison University, Harrisonburg, VA, USA}
\author{I.~Niculescu}
\affiliation{James Madison University, Harrisonburg, VA, USA}
\author{Y.~Okayasu}
\affiliation{Tohoku University, Sendai, Japan}
\author{A.~K.~Opper}
\affiliation{Ohio University, Athens, OH, USA}
\author{C.~Perdrisat}
\affiliation{College of William and Mary, Williamsburg, VA, USA}
\author{V.~Punjabi}
\affiliation{Norfolk State University, Norfolk, VA, USA}
\author{A.~W.~Rauf}
\affiliation{University of Manitoba, Winnipeg, Manitoba, Canada}
\author{V.~M.~Rodriquez}
\affiliation{University of Houston, Houston, TX, USA}
\author{D.~Rohe}
\affiliation{Basel University, Basel, Switzerland}
\author{J.~Seely}
\affiliation{Laboratory for Nuclear Science, Massachusetts Institute of Technology, 
  Cambridge, MA, USA}
\author{E.~Segbefia}
\affiliation{Hampton University, Hampton, VA, USA}
\author{G.~R.~Smith}
\affiliation{Physics Division, TJNAF, Newport News, VA, USA}
\author{M.~Sumihama}
\affiliation{Tohoku University, Sendai, Japan}
\author{V.~Tadevosyan}
\affiliation{Yerevan Physics Institute, Armenia}
\author{L.~Tang}
\affiliation{Hampton University, Hampton, VA, USA}
\affiliation{Physics Division, TJNAF, Newport News, VA, USA}
\author{A.~Villano}
\affiliation{Rensselear Polytechnic Institute, Troy, NY, USA}
\author{W.~F.~Vulcan}
\affiliation{Physics Division, TJNAF, Newport News, VA, USA}
\author{F.~R.~Wesselmann}
\affiliation{Norfolk State University, Norfolk, VA, USA}
\author{S.~A.~Wood}
\affiliation{Physics Division, TJNAF, Newport News, VA, USA}
\author{L.~Yuan}
\affiliation{Hampton University, Hampton, VA, USA}
\author{X.~Zheng}
\affiliation{Argonne National Laboratory, Argonne, IL, USA}
    \date{\today}

\begin{abstract}
Cross sections for the p($e,e'\pi^{+}$)n process on $^1$H,
$^2$H, $^{12}$C, $^{27}$Al, $^{63}$Cu and $^{197}$Au targets were 
measured at the Thomas Jefferson National Accelerator Facility 
(Jefferson Lab) in order to extract the nuclear transparencies. Data 
were taken for four-momentum transfers ranging from $Q^2$=1.1 to 4.8 
GeV$^2$ for a fixed center of mass energy of $W$=2.14 GeV. The ratio 
of $\sigma_L$ and $\sigma_T$ was extracted from the measured cross 
sections for $^1$H, $^2$H, $^{12}$C and $^{63}$Cu targets at 
$Q^2$ = 2.15 and 4.0 GeV$^2$ allowing for additional studies of the 
reaction mechanism. The experimental setup and the analysis of the data 
are described in detail including systematic studies needed to obtain 
the results. The results for the nuclear transparency and the 
differential cross sections as a function of the pion momentum at the 
different values of $Q^2$ are presented. Global features of the data are 
discussed and the data are compared with the results of model calculations for 
the p($e,e'\pi^{+}$)n reaction from nuclear targets.
\end{abstract}


\maketitle

\section{Introduction}
\label{sec:intro}

A fundamental challenge in nuclear physics is trying to understand the 
structure of hadrons in terms of their quark-gluon constituents, which 
are governed by the underlying theory of strong interaction called Quantum Chromodynamics (QCD). Measurements of exclusive processes such as pion electroproduction make it possible to extract meson form factors and study the quark-gluon distributions in the nucleon. However, in order to develop a description of the 
atomic nucleus based on QCD, one also needs to understand how the properties 
and interactions of hadrons change in the nuclear medium. Measurements of exclusive processes in the nuclear medium sometimes help one to understand the interactions because QCD has definite predictions for exclusive processes  in the medium. One such prediction is the phenomenon of color transparency (CT). 

 In the study of CT, we can observe the interplay between the creation of quark systems with small transverse size, as we would expect from QCD, and the possible suppression of interactions of such point-like configurations (PLC) with the nuclear medium. A further complication is introduced by the formation time
of the final hadron, which limits the lifetime of the PLC. To understand
this phenomenon, it is thus important to study the dependencies on the 
four-momentum transfer squared of the virtual photon, $Q^2$ (size of the
initial PLC, typically scales as $\sim 1/Q$), hadron momentum (formation length), and target mass $A$ (path length through the medium). 

Due to the higher probability to form a PLC, a two-quark rather than a 
three-quark system is preferable, making pion electroproduction a natural 
choice for an initial study. However, pion electroproduction introduces the assumption of a quasi-free interaction of the photon with a virtual pion in the medium. The plausibility of this assumption can be addressed by comparing the ratio of the longitudinal to transverse cross sections from a nucleus with those obtained using a nucleon target, as differences in the behavior of this ratio would indicate effects of the nuclear medium resulting in the breakdown of the quasi-free assumption. 

Nuclear transparency is a natural  observable in the quest to identify the onset of color transparency. Nuclear transparency is defined as the ratio of the cross section per nucleon for a process on a bound nucleon inside a nucleus to that from a free nucleon.  Since the nuclear medium is not opaque to hadrons, color transparency will lead to an increase in nuclear transparency as a function of momentum transfer, hadron momentum and target mass. This simultaneous increase of nuclear transparency is distinct from those from other conventional processes such as rescattering. Moreover, it illustrates the need for careful experiment design that simultaneously measures the $Q^2$ and $A$ dependence of nuclear transparency and at the same time tests alternative reaction mechanisms that can produce potential increases in nuclear transparency. 

With the availability of high-intensity, continuous electron beams up to 
6 GeV at Jefferson Lab it became possible for the first time to 
determine simultaneously the $A$ and $Q^2$ dependences of the differential 
pion cross sections for $Q^2$ in the 1-5 GeV$^2$ range. 
These data were acquired in 2004 for $^1$H, $^2$H, $^{12}$C, $^{27}$Al, 
$^{63}$Cu and $^{197}$Au targets. 
The $A$ and $Q^2$ dependence of the extracted transparency were 
published in~\cite{Clasie}.

The purpose of this work is to describe the experiment and analysis and
to present and discuss additional results. The article is organized as 
follows: the next sections describe methods for extracting 
nuclear transparencies and discuss some earlier data. 
Section~\ref{sec:formalism} introduces the basic cross section formalism 
of pion electroproduction. 
In section~\ref{sec:experiment} the experiment 
is described including the experimental setup. The data analysis is 
described in section~\ref{sec:analysis}. The determination of the
cross sections and the extraction of the nuclear transparencies and their
associated uncertainties are described in section~\ref{sec:dettransp} and
section~\ref{Sec:sys}. The results are presented in 
section~\ref{sec:results}. The global features of the cross section ratio 
and the nuclear transparencies are discussed and a comparison is made with 
results of recent calculations.

\section{Methods of determining nuclear transparency from data}
\label{sec:transpmethods}

In a simple picture of pion electroproduction from the nucleus,
the electron exchanges a virtual photon with a proton which is moving 
due to its Fermi momentum. The struck proton turns into a neutron 
ejecting a positively charged pion (quasi-free approximation). In 
the quasi-free picture, the ejected pion may interact with the residual 
nucleons and the fraction of pions which can escape from the nucleus 
is the nuclear transparency of pions. In reality, the scattering process 
is more complicated and the
deviations from this picture reveal much about the nucleus and their
constituents. In the simple quasi-free picture, the ratio of the 
longitudinal to transverse cross section from a bound  proton inside 
the nucleus is expected to be the same as that from a free proton.

Assuming the dominance of the quasi-free process, one can extract 
the nuclear transparency  of the pions by taking the ratio of the acceptance
corrected cross sections from the nuclear target to those from the
proton. Nuclear transparency quantifies the interactions of the
outgoing pion with the residual nucleons on its way out of the
nuclei, and is thus the key observable in searching for CT effects. 
Color Transparency was first proposed  by Brodsky 
and Mueller~\cite{BroMue} in 1982, and refers to the vanishing of the 
hadron-nucleon interaction for hadrons produced in exclusive processes. 
At sufficiently high momentum transfers, where the hadron is produced 
with small transverse size, $b_{\perp} \approx 1/Q$~\cite{BM}, 
the fast-moving hadron also has a compact longitudinal size due to 
Lorentz contraction. Such a reduced size  quark-gluon state is 
called a point-like configuration (PLC). The  formation length, the 
distance over which the PLC travels before the hadron reaches its final 
size can be written,
\begin{equation}
  l_{f} \approx \frac{ \beta \cdot T_{lifetime} }{\sqrt{1 - \beta^2}}
\end{equation}

assuming a linear expansion in time, where $\beta$ is the speed of the hadron 
in the lab frame and $T_{lifetime}$ is the life time of the PLC in its rest 
frame. 

QCD predicts that the cross section for 
the interaction of a small $q\bar{q}$ dipole is proportional to $b^2$ 
in the leading order approximation, where $b$ is the transverse separation 
between the $q$ and the $\bar{q}$. 

If the $q\bar{q}$ dipole remains small (pointlike) over the range of the 
nuclear system, the nucleus will be transparent to the produced hadron. This 
can be accomplished at large momentum transfers where the formation length is 
sufficiently large.

At finite energies, however, the mechanism for the 
expansion and contraction of the interacting small system is more complex.
The quantum diffusion model~\cite{Farrar} assumes that quarks 
separate  in the transverse direction at the speed of light and that their 
quark separation is proportional to 
$\sqrt{z}$, where $z$ is the longitudinal distance from the production 
point to the position of the particle. The formation length in this 
model is determined from the average value of the dominant energy 
denominator.
\begin{equation}
  l_{f} \approx 2P_{h} \left \langle \frac{1}{M_n^2-M_h^2} \right \rangle
\end{equation}
where $P_h$ and $M_h$ are the momentum and mass of the hadron,
respectively, while $M_n$ is the mass of a typical intermediate
state of the hadron. 

The precise value for $\Delta M^2= (M_n^2-M_h^2)^{-1}$ is a matter of 
great uncertainty~\cite{Farrar, Larson06ge} with estimates between 0.25 and 1.4 GeV$^2$.

The onset of color transparency is expected at lower energies for production of
mesons, for instance of pions, than for production of baryons, because 
the quark-antiquark pair is more likely to form a small size 
object~\cite{blattel}. 

In pion electroproduction, the coherence length, defined as the distance over 
which the virtual photon fluctuates into a $q\bar{q}$ pair, is smaller than the nucleon radius at JLab energies, and for the present experiment it is essentially constant. This removes any coherence length dependence through t-channel $\pi-\rho$ exchange, in which the incident electron scatters from a virtual photon
bringing it on-shell. Such a coherence length dependence can mimic a CT-like energy dependence.

Color transparency has recently been put into the context of a QCD
factorization theorem for longitudinally polarized photons 
in meson electroproduction. It predicts that for sufficiently 
high values of $Q^2$, at fixed $x_B$ and fixed momentum transfer to the 
nucleon, $-t$, the amplitude can be expressed in terms of a 
short-distance (hard) process, a distribution amplitude describing the 
final state meson, and Generalized Parton Distributions 
(GPDs)~\cite{Collins}. The latter describe the long-distance (soft) physics.

The factorization theorem has been proven at asymptotically high $Q^2$, but 
showing its validity at finite, high $Q^2$ requires stringent tests of all
necessary conditions.

The unambiguous observation of the onset of CT is a critical precondition for
the validity of the factorization theorem. This is because where CT applies
the outgoing meson retains a small transverse size (inter-quark distance)
while soft interactions like multiple gluon exchange between the meson
produced from the hard interaction and the baryon are highly suppressed. 
Factorization is thus rigorously not possible without CT.

\section{Early extractions of nuclear transparency}
\label{sec:prev}

The first experiments designed to search for color transparency used the 
$^{12}C(p,2p)$ reaction and were performed at Brookhaven National Laboratory
(BNL)~\cite{Carroll,Mardor,Leksa}. The nuclear transparency was defined as the
ratio of the elastic $p-p$ cross section in the nucleus and hydrogen. This 
ratio initially increased as a function of the beam energy and then decreased, 
with a peak near 9 GeV. Though this behavior was not predicted by traditional 
nuclear physics calculations, it is currently not attributed to CT. Instead it  
is associated with nuclear filtering~\cite{Ralston,Pire}, 
which is related to the suppression of the long range components in the wave 
function, or to the threshold for charm resonance production~\cite{Stan3}.

The nuclear transparency was also measured using the A($e,e^\prime p$) reaction 
at Stanford Linear Accelerator Center (SLAC)~\cite{Makins,Neill} and at 
Jefferson Lab~\cite{Garrow,Abbott}. In this case, the nuclear transparency was defined as the ratio of the experimental
yield and the one obtained from Plane Wave Impulse Approximation.
The ratio was found to be energy independent from $Q^2$ $\approx$ 2 GeV$^2$ to 
8.1 GeV$^2$ for deuterium, carbon, iron, and gold targets, thus indicating no 
significant CT effect. The absence of the CT effect in this channel has been 
interpreted as an indication that the proton formation length may only have 
been as large as internucleonic distances, rather than the size of the 
nucleus in these experiments~\cite{Laget}.

CT measurements using coherent and incoherent $\rho^0$ production have been
performed at Fermilab~\cite{Adams} and more recently at DESY~\cite{Airape}.
The nuclear transparency was defined as $T=\sigma_A/A \sigma_H$. The results
from Fermilab were parametrized with a function $T=A^{\alpha-1}$, and at a 
first glance, the observed positive slope of $\alpha$ as a function of $Q^2$ 
appeared to contradict the flat $Q^2$ dependence predicted by the Glauber 
multiple scattering mechanism. However, the results have since been 
interpreted as a coherence length effect~\cite{Acker}, in which for kinematics 
with large coherence length, the virtual $q\bar{q}$ pair may undergo 
interactions with the nucleus before the hard interaction that puts it 
on-shell. The kinematics of the Fermilab experiment were not at a 
constant $l_c$, and thus the variation of $\alpha$ with $Q^2$ was attributed 
to a reduction in the initial state interactions rather than to a reduction in 
final state interactions. The more recent measurements from DESY, ranging 
between $Q^2$=0.9-3 GeV$^2$ and at constant $l_c$, showed a rise in nuclear 
transparency with $Q^2$ consistent with theoretical calculations of CT~\cite{Airape}. 

The most convincing evidence for the existence of color transparency comes
from an experiment performed at Fermilab~\cite{Aitala}. There, the cross 
section of diffractive dissociation of 500 GeV/$c$ pions into dijets was 
measured and parametrized with $\sigma$=$\sigma_0 A^\alpha$, where 
$\sigma_0$ is the $\pi-N$ cross section in free space. The free parameter, 
$\alpha$, was fit to the data with the result $\alpha \sim$1.6. This result 
was in agreement with calculations assuming 100\% color transparency and was 
very different from the normal $\pi-N$ cross section, which has a dependence 
$\sigma$=$\sigma_0 A^{2/3}$.

At low energies, hints of CT effects have also been observed from pion 
photoproduction from helium, $^4He(\gamma,\pi^-p)$, at Jefferson Lab~\cite{Dutta}. CT can be measured in these reactions where $Q^2$=0 by measuring the cross section vs. the four momentum transfer squared, $t$ to the hadron system. The data
showed 2$\sigma$ deviations from the traditional Glauber calculations and the 
slope of the data vs. $-t$ was in better agreement with calculations including
CT.

\section{Elementary pion electroproduction}
\label{sec:formalism}

\subsection{Kinematics}
\label{sec:kinematics}

\begin{figure}[t]
\begin{center}
 \includegraphics[width=9cm]{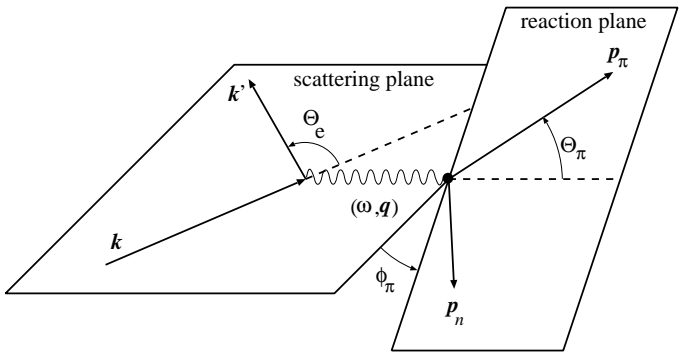}
 \caption{\it Kinematics of the p($e,e^\prime,\pi^+$)n reaction in the laboratory frame.}
 \label{fig:eepi_kinematics}
\end{center}
\end{figure}

The kinematics of the pion electroproduction reaction studied here are 
shown in Figure~\ref{fig:eepi_kinematics}. The incident electron has a 
three momentum of $\mathbf{k}$. The scattered electron has a three momentum 
$\mathbf{k}'$ and travels at a polar angle $\theta_e$ in the laboratory
frame with respect to the direction of the incident beam. The three
momentum vectors of the incoming and outgoing electron define the 
scattering plane. The corresponding four-momenta are 
$k \equiv (E,\mathbf{k})$ and $k'\equiv (E',\mathbf{k'})$. The virtual
photon carries a four-momentum $q \equiv (\omega,\mathbf{q})$, which 
is given by $q \equiv k-k'$. The reaction plane is defined by the 
three-momentum $\mathbf{q}$ and the three-momentum vector of the 
produced pion $\mathbf{P}_\pi$ and makes an angle $\phi_{\pi}$ with respect
to the scattering plane. The angle (in the lab system) between 
$\mathbf{p}_\pi$ and $\mathbf{q}$ is $\theta_\pi$.

The pion electroproduction reaction can be described using 
three Lorentz~invariants. In addition to $Q^2$, we use the invariant mass
of the virtual photon-nucleon system, $W$, which can be expressed as
$W=\sqrt{M_p^2+2 M_p \omega-Q^2}$, where $M_p$ is the proton mass,
and $t=(P_\pi-q)^2$ is the Mandelstam variable.

\subsection{Cross sections}
\label{sec:cross_sections}

The pion electroproduction cross section from a stationary
proton in the one photon-exchange approximation is~\cite{Nozawa}:
\begin{equation}
\label{eq:fivefoldxs}
  \frac{d^5\sigma}{d\Omega_{e'}dE_{e'}d\Omega_{\pi}}=\Gamma_{\nu}\frac{d^2\sigma}{d\Omega_{\pi}} 
\end{equation} 
where 
\begin{equation}
\label{eq:gammanu}
\Gamma_{\nu} =
    \frac{\alpha}{2\pi^2}\frac{E_{e'}}{E_e}\frac{K_{eq}}{Q^2}\frac{1}{1-\epsilon}
\end{equation} 
is the virtual photon flux, where $\alpha$ is the fine structure 
constant. The factor $K_{eq} = (W^2-M_p^2)/(2M_p)$ is the equivalent 
photon energy, and
\begin{equation}
\label{eq:epsilon}
   \epsilon = (1+\frac{2|{\bf q}|^2}{Q^2}\tan^2\frac{\theta_e}{2})^{-1} 
\end{equation} 
is the longitudinal polarization of the virtual photon.
The two-fold differential cross section for a stationary proton 
target can be written as
\begin{equation}
\label{eq:d2sigma}
    \frac{d^2\sigma}{d\Omega_{\pi}}= J \frac{d^2 \sigma}{dt d\phi},
\end{equation} 
where the solid angle of the pion, $\Omega_{\pi}$ is determined in 
the lab frame, and $J$ is the Jacobian for the transformation 
from $\Omega_\pi$ to $t,\phi$. The two-fold cross section can be 
separated into four structure functions, which correspond to the 
polarization states of the virtual photon, a longitudinal (L), a 
transverse(T), and two interference terms (LT and TT):
\begin{eqnarray}
\label{eq:sepsig1}
2\pi \frac{d^2 \sigma}{dt d\phi} & = & 
   \epsilon \hspace{0.5mm} \frac{d\sigma_{\mathrm{L}}}{dt} +
   \frac{d\sigma_{\mathrm{T}}}{dt} + \sqrt{2\epsilon (\epsilon +1)}
   \hspace{1mm}\frac{d\sigma_{\mathrm{LT}}}{dt}
   \cos{\phi}  \nonumber \\
   & & + \epsilon \hspace{0.5mm}
   \frac{d\sigma_{\mathrm{TT}}}{dt} \hspace{0.5mm} \cos{2 \phi} ,
 \end{eqnarray}
The interference terms vanish in parallel kinematics ($\theta_{\pi}=0$) 
because of their dependence on $\theta_{\pi}$~\cite{sintheta}.

The four structure functions can be separated if measurements at different 
values of $\epsilon$ and $\phi_{\pi}$ are performed (L/T separation), 
where $W, Q^{2}$ and $t$ are kept constant. The photon polarization 
$\epsilon$ can be varied by changing the electron energy and 
scattering angle.
 \begin{table}[!htb]
 \centering  
 \renewcommand{\arraystretch}{1.2}
 \begin{tabular}{cccccccccc} \hline
  $Q^2$ & $W$  & $-t_{min}$ & $E$ & $\theta_e$ & $E^\prime$ & $\theta_{\pi}$ & $P_\pi$ & $k_\pi$ & $\epsilon$ \\
  GeV$^2$ & GeV & GeV$^2$ & GeV & deg & GeV & deg & GeV/$c$ & GeV/$c$ & \\
         \hline
1.10 & 2.26 & 0.050 & 4.021 & 27.76 & 1.190 & 10.61 & 2.793 & 0.23 & 0.50\\
2.15 & 2.21 & 0.158 & 5.012 & 28.85 & 1.730 & 13.44 & 3.187 & 0.41 & 0.56\\
3.00 & 2.14 & 0.289 & 5.012 & 37.77 & 1.430 & 12.74 & 3.418 & 0.56 & 0.45\\
3.91 & 2.26 & 0.413 & 5.767 & 40.38 & 1.423 & 11.53 & 4.077 & 0.70 & 0.39\\
4.69 & 2.25 & 0.527 & 5.767 & 52.67 & 1.034 & 10.63 & 4.412 & 0.79 & 0.26\\
2.16\footnote{low $\epsilon$ setting used for L-T separations} & 2.21 & 0.164 & 4.021 & 50.76 & 0.730 & 10.60 & 3.187 & 0.42 & 0.27\\
4.01\footnote{low $\epsilon$ setting used for L-T separations} & 2.14 & 0.441 & 5.012 & 55.88 & 0.910 & 10.55 & 3.857 & 0.71 & 0.25\\
2.16\footnote{test point to check the dependence on $W$ and $k_\pi$} & 1.74 & 0.374 & 4.021 & 32.32 & 1.730 & 19.99 & 2.074 & 0.65 & 0.63\\
         \hline	   
 \end{tabular}
 \caption{\label{table:kine} 
\it The central kinematic settings for the pion transparency experiment, $\theta_q$ is the angle between the virtual photon 3-momentum ${\bf q}$ and the beam direction in the lab frame, $k_\pi$ is the magnitude of the three momentum of the virtual struck pion in the quasi-free knockout picture, and $x=Q^2/(2M\nu)$. 
}
 \end{table}

For nuclei, where there is a new degree of freedom due to the Fermi
momentum of the struck nucleon, the differential pion electroproduction
cross section is given by,
\begin{equation}
\label{eq:sixfoldxs}
  \frac{d^6\sigma}{d\Omega_{e'}dE_{e'}d\Omega_{\pi}dP_{\pi}}=\Gamma_{\nu}\frac{d^3\sigma}{d\Omega_{\pi}dP_{\pi}},
\end{equation}
The three-fold differential cross section, $\frac{d^3\sigma}{d\Omega_{\pi}dP_{\pi}}$ can be separated into longitudinal, transverse and interference terms just as in Eqns.~\ref{eq:d2sigma},\ref{eq:sepsig1}. 

\begin{figure*} 
\begin{center}
\resizebox{5.8in}{3.4in}{\includegraphics{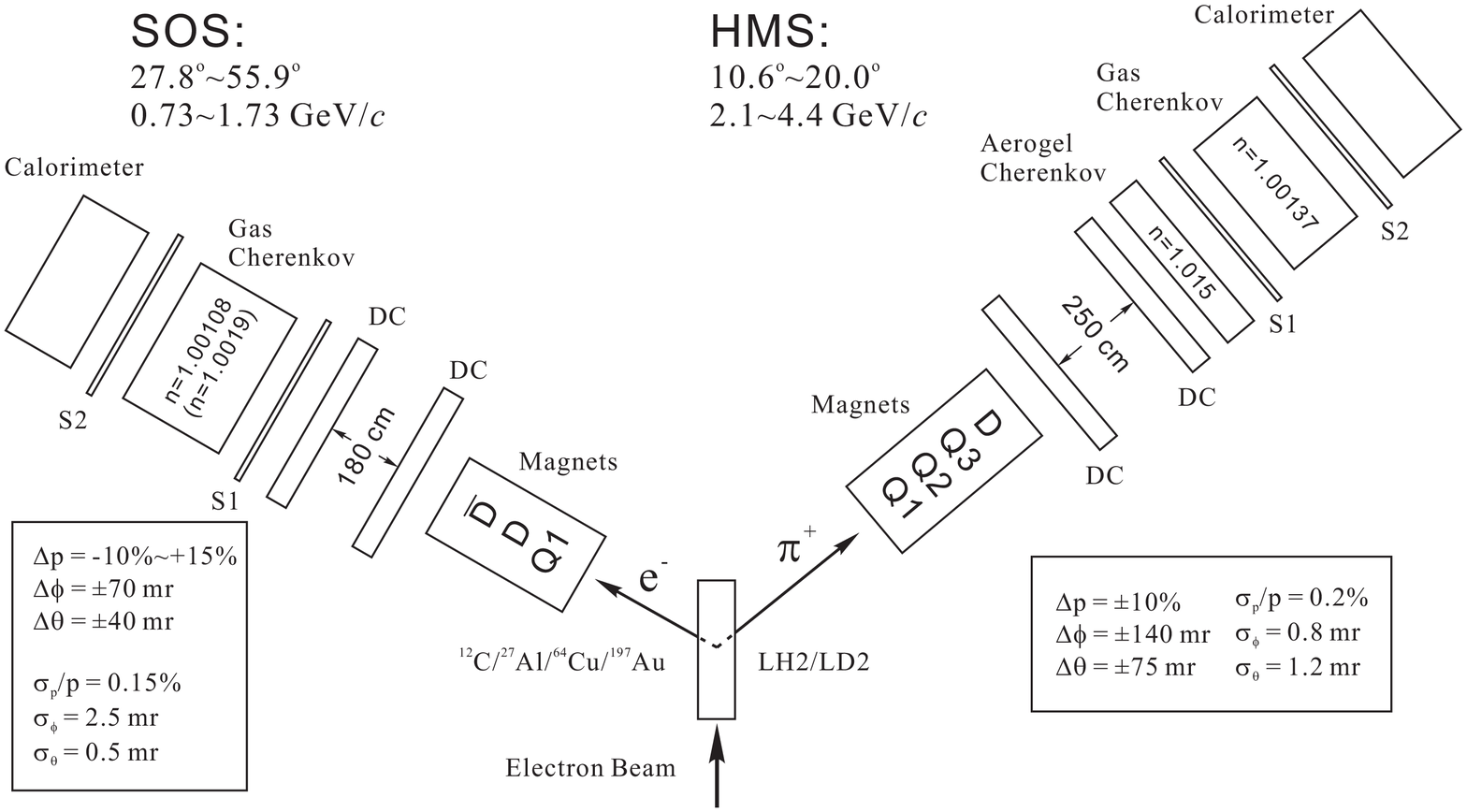}}
\caption{\it Schematic view of the experimental setup for E01-107.} 
\label{fig:over}
\end{center} 
\end{figure*}

\section{Experiment and Setup}
\label{sec:experiment}

The E01-107 experiment (pionCT) was carried out in Hall C at 
Jefferson Lab~\cite{JLab} in 2004. A schematic view of the experimental 
setup is shown in Fig.~\ref{fig:over} and the kinematic settings of the 
measurements are shown in Table~\ref{table:kine}. 
The continuous wave (100\% duty factor) electron beam from the CEBAF 
accelerator has a characteristic 2 ns micro-structure that arises from the 
1.5GHz rf structure of the accelerator and the 3 hall beam splitting scheme. This 2 ns structure is used in the analysis to identify ($e,e^\prime \pi^+$) coincident events. The electron beam with 
energies of up to 5.8 GeV was incident on 
liquid hydrogen and deuterium, and solid foil targets of $^{12}$C, 
$^{27}$Al, $^{63}$Cu and $^{197}$Au. For the cryotargets, a 4.0~cm diameter
cylindrical cell with an axis perpendicular to the beam direction was
used. The cell walls were made from an aluminum alloy with a thickness of
0.01 cm. The integrated beam current (35-80 $\mu$A) was measured using a pair 
of microwave cavities with a DC current transformer to an accuracy of 0.4\%. 
To reduce target density fluctuations in the cryotargets arising from beam 
heating, the beam was rastered to a 2$\times$2 mm$^2$ profile. No target
density fluctuations were measured to an accuracy of 0.6\%.

The scattered electrons were detected in the short orbit spectrometer
(SOS). The SOS consists of three non-superconducting magnets, one 
quadrupole followed by two dipoles, which share a common yoke. 
The quadrupole focuses in the non-dispersive direction, the first 
dipole bends particles with the central momentum upwards by 33$^\circ$, 
and the second one bends them downwards by 15$^\circ$.
The SOS spectrometer design was
optimized for detection of unstable and short-lived particles and
thus has a flight path of only ~10 m. The electroproduced pions were 
detected in the high momentum spectrometer (HMS). The HMS consists of 
three superconducting quadrupole magnets followed by a momentum analyzing
vertical-bend superconducting dipole used in a point-to-point tune for 
the central ray. The detector package is mounted
near the focal plane of the optical system, which is located inside a 
concrete shield house. The HMS has a 26~m 
pathlength and a maximum central momentum of 7.3~GeV/$c$
A detailed 
description of the spectrometers and the spectrometer hardware is given 
in~\cite{arr98,dutta03}. Selected properties of the two spectrometers are listed 
in Table~\ref{tab:HMSspec}.
 \begin{table}[!htb]
     \centering  
     \renewcommand{\arraystretch}{1.2}
\small
     \begin{tabular}{||l|l|l||} \hline       
        Quantity &  HMS & SOS \\
        \hline
        Max. Central Mom. & 7.3~GeV/$c$\footnote{To date, the HMS has been operated to 6.0~GeV/$c$} & 1.74~GeV/$c$ \\
        Optical Length & 26.0~m & 7.4~m\\
        Momentum Acceptance & $\pm$10\% & $\pm$20\% \\
        Solid Angle\footnote{The solid angle and angular acceptances 
        are given for the large collimators in both 
        the HMS and SOS spectrometers.} & 6.7~msr & 7.5~msr \\
        In-Plane Ang. Acc.$^a$ & $\pm$27.5~mrad & $\pm$57.5~mrad \\
        Out-of-Plane Ang. Acc.$^a$ & $\pm$70~mrad & $\pm$37.5~mrad \\
        \hline  
     \end{tabular}
     \caption{\label{tab:HMSspec} \it Selected properties for the HMS and SOS.}
 \end{table}

Both spectrometers are equipped with multiwire drift chambers for
particle tracking, and segmented scintillator hodoscope arrays for 
time-of-flight measurements and triggering. 

The HMS has a gas \v{C}erenkov and a lead glass calorimeter in the 
detector stack for $K^+$/$\pi^+$ separation. For this experiment, the 
\v{C}erenkov was filled with $C_4F_{10}$ gas at 97~kPa. The corresponding index
of refraction is 1.00137 resulting in momentum thresholds of 2.65 GeV/$c$ 
for $\pi^+$ and 9.4 GeV/$c$ for $K^+$. 
An aerogel \v{C}erenkov 
detector~\cite{aero} was also used in the detector stack. It was 
used to identify $\pi^+$ for central momentum settings below 3.1 GeV/c. The
aerogel had an index of refraction of 1.015 giving thresholds of 0.8 GeV/$c$
for pions and 2.85 GeV/$c$ for kaons.

The SOS has a combination of a gas \v{C}erenkov and a lead-glass calorimeter 
for $e^-/\pi^-$ separation. The nominal SOS \v{C}erenkov detector is filled
with $CCl_2F_2$ gas at 101~kPa with an index of refraction of 1.00108. During part of the experiment ($E_e$ =5.767 GeV) the nominal SOS \v{C}erenkov detector was replaced by one filled with $C_4F_{10}$ gas at 143~kPa with an index of refraction of 1.0019. The corresponding thresholds for electrons were below
10 MeV/$c$ and those for pions were 3 GeV/$c$ (nominal) or 2.27 GeV/$c$ (new). 

\section{Data Analysis}
\label{sec:analysis}

The raw data were processed and combined with additional information
from the experiment like momentum and angle settings of the 
spectrometers, detector positions, and beam energy to give the 
particle trajectories, momenta, velocities, energy deposition, and to 
perform particle identification.

\subsection{Event Reconstruction}
\label{sec:recon}

The spectrometer quantities $x$, $y$, $x^\prime$ and $y^\prime$, were
deduced from
reconstruction of the wire chamber data. These quantities are the
vertical and horizontal positions of the track at the midpoint
between the wire chambers, and the gradients of that track with
respect to the spectrometer central ray. The target quantities
$y_{tar}$, $x'_{tar}$, $y'_{tar}$ and $\delta$ were determined from
the spectrometer quantities by suitable transformation
functions. These quantities are the horizontal position of the event,
the horizontal and vertical gradients of the track with respect to the
spectrometer central ray and the momentum of the particle given as a
percentage above the central momentum setting of the spectrometer,
respectively. A special data set, using a series of foils
targets placed at well determined positions in the target region was
taken to calibrate the transformation matrix. 

\subsection{Particle Identification and event selection}
\label{sec:pid}

Electrons were selected with a SOS gas \v{C}erenkov cut of 
N$_{photoelectrons}$ $>$ 1.0 for the nominal or N$_{photoelectrons}$ $>$ 5.0 for 
the new detector. The efficiency of the cut was determined, for each of the SOS central momentum settings, using a sample of electrons that was identified using the lead-glass calorimeter. The resulting efficiency was always higher than 99.2\% with an uncertainty of 0.2\%. The pion rejection ratio was 100:1 for the 
nominal and 300:1 for the new detector. 
In the HMS, the aerogel and gas \v{C}erenkov detectors were used to select  
$\pi^+$. The aerogel was used for additional particle identification when 
the central momentum setting of the HMS, $P_{HMS}$, was less than 3.2 GeV/$c$.
The lower limit of the aerogel efficiency was determined using tight cuts on 
the coincidence time, and was found to be $>$98.8$\pm$ 0.5\% for a threshold
cut of N$_{photoelectrons}>$0.7. The rejection ratio for the HMS Aerogel 
\v{C}erenkov detector was approximately 5:1 at $P_{central}$ = 2.1 and 2.8 
GeV/$c$. 
A cut on the gas \v{C}erenkov detector varying between N$_{photoelectrons}>$0.7 
and N$_{photoelectrons}>$2.0 was used when $P_{HMS}\geq 3.2$ GeV/$c$. 
The corresponding cut efficiencies were determined using tight cuts on the
coincidence time and the aerogel \v{C}erenkov detector to remove protons, and 
were found to be $>$ 98.2$\pm$0.5\%  
The cut efficiency was parametrized as a function of the HMS fractional 
momentum, $\delta_{HMS}$ to take into account that pions at negative 
$\delta_{HMS}$ are closer to the momentum threshold of the detector compared 
to those at positive $\delta_{HMS}$. The resulting rejection ratio was 50:1 at 
$P_{central}$=3.2 GeV/c and 300:1 at $P_{central}$ = 4.4 GeV/c with an uncertainty of 0.2\%. 

\subsection{Efficiencies}
\label{sec:efficiencies}

The raw yield was normalized by the beam charge, 
and the data were corrected for detector efficiencies, computer
dead time, correction to target thickness and nuclear absorption in
the detection materials.

\subsubsection{Tracking efficiency}
\label{sec:tracking_eff}

The tracking efficiency is the probability of finding a track from 
experimental signals from the wire chambers when a charged particle passes 
through them. It depends on the efficiency of the wire chambers and also on 
the algorithm used to construct a track for a given event. 
The tracking efficiency was $>$92\% for the HMS and $>$96\% for the SOS, and was calculated run-by-run for this experiment. The uncertainties to the tracking efficiencies are 
estimated from the stability of the tracking efficiency as a function of
the total rate in the spectrometer, and were found to be 1.0 \% for 
the HMS and 0.5\% for the SOS. The rate dependence of the tracking efficiency, 
which is related to the increased probability to have multiple tracks at high
rates was taken into account with a tracking efficiency calculation including
multiple track events~\cite{pionfflong,vladis}.

\subsubsection{Trigger Efficiency}
\label{subsect:triggereff}

The HMS and the SOS spectrometers each contain four layers of scintillators 
and a requirement for a single-arm trigger is that 3 out of 4 layers must have 
a hit. The 3-out-of-4 efficiency was found to be above 99.5\% for all runs in
both spectrometers with an uncertainty of 0.5\% for the HMS and 0.5\% for the SOS.

\subsubsection{Computer and Electronics dead times}
\label{subsec:deadtimes}

The computer dead time, which are due to the finite time required by the computers to process an event, can be directly calculated from the number of generated pretriggers and the number of accepted triggers. The computer dead time during the pionCT experiment was about 25\%. The electronics dead time, due to the finite time required by the electronic modules was estimated from copies of the original pretrigger signal at varying gate widths. The correction due
to electronics dead time was less than 1\%. The total uncertainty (0.2-0.5\%) 
due to the dead time corrections was dominated by the computer dead time.

\subsubsection{Coincidence Blocking}
{\label{sect:coinblock}}

Coincidence events can be blocked when a non-coincident event arrives in one
of the spectrometers just before the coincident event. The ``coincidence 
blocking'' events are lost from the data due to the coincidence time cuts in 
the analysis. The coincidence blocking correction was estimated from the
rate dependence of the number of blocked events, and was found to be less than
0.7\% with an uncertainty of about 0.2\%.

\subsubsection{Pion Absorption}
{\label{sect:piabsorb}}

Pions may interact through the strong nuclear force with the nucleus, for 
instance in the target material, the window of the scattering chamber, or the 
windows of the spectrometer. These events are lost before reaching the detectors
in the HMS detector hut. We account for these events by correcting for pion absorption in the HMS. The transmission for pions in the HMS is about 
95\%~\cite{pionfflong} and depends weakly on the pion momentum between 2.1 and 4.4 GeV/c. In addition some pions interact with the scintillator material producing events that reconstruct at lower pion velocity ($\beta_{\pi}$) or even $\beta_{\pi}=0$ and lower coincidence time. Using a 2-dimensional cut on $\beta_{\pi}$ and coincidence time, such events were included in our pion yield. The efficiency of such a cut was studied using $(e,e'\pi^-)$~\cite{tanyathesis} and $H(e,e'p)$ ~\cite{tanyathesis,derekthesis} data. The difference between the 2-dimensional cut and a simple cut on pion velocity that does not include the pion with lower velocity and the $\beta_{\pi}=$0 events was found to be within the uncertainty associated with the HMS detector efficiency and the uncertainty of the pion absorption correction. The uncertainty in the pion absorption correction was estimated from the difference between the calculated pion 
transmission and the measured proton transmission, and was found to be 2\% in 
the absolute cross section determination~\cite{blok08,pionfflong}. The difference in target thickness leads to an additional $A$ dependent uncertainty which is estimated to be 0.5\%.

\subsubsection{Backgrounds}
{\label{sect:backgrounds}}

The coincidence time was calculated from the time difference between the 
SOS and HMS triggers, and was used to help identify $e^-$/$\pi^+$ coincidences from $e^-$/$p$ coincidences. Corrections to the coincidence time include the path length of the tracks through the magnetic elements of the spectrometer, difference in signal propagation times in cables, pulse height corrections for the signals from the scintillators, and subtraction 
of the time required for the light to travel in the scintillators from the 
event position to the PMT. Random coincidences, resulting from an electron 
and pion from different beam buckets, have a coincidence time that is offset 
from the in-time by multiples of 2 ns. The in-time 
$e^-$/$\pi^+$ events were selected with a cut on the coincidence time 
around the central peak. 
The random background was estimated by averaging over three bunches, that could be readily identified by eye, to the right and three bunches to the left of the in-time peak, and then 
subtracting it from the in-time yield.

Background from the cryotarget cell walls, less than a few percent, was 
measured and subtracted using an aluminum target of approximately seven 
times the thickness of the target cell walls. The contribution of the cell
walls was small (less than 5\%), and due to the high statistical accuracy of 
the dummy target data, the contribution of the subtraction was $<$ 0.1\% to 
the total uncertainty.

\subsubsection{Missing mass}
{\label{sect:missmass}}

Once true $e^-$/$\pi^+$ coincidences are identified, the missing mass
of the recoiling nucleon system was reconstructed from the measured
quantities. In the present analysis, a cut in the missing mass was 
used to ensure that no additional pions were produced in the case of 
hydrogen, and to minimize the contribution of multi-pion events to less
than 5\% with an uncertainty of $<$ 0.4\% (see section~\ref{sec:multipion} 
for a more detailed discussion). 

\section{Determination of the nuclear transparency}
\label{sec:dettransp}

To extract cross section information from the data, 
the measured yields were compared to the results of a Monte Carlo 
simulation for the actual experimental setup 
(see section~\ref{sec:MC}), which included a realistic model of 
the pion electroproduction cross section. When the model describes the
dependence on $W$, $Q^2$, $t$, and $\theta_\pi$ ($P_{\pi}$) of the 
four structure function in equation~\ref{eq:sepsig1} well, the 
cross section for any value of $W$ and $Q^2$ in the acceptance can
be determined as
\begin{equation}
\label{eq:sigextract}
\left(\frac{d\sigma}{dt}\right)_{data} = \frac{Y_{data}}{Y_{mc}} \left(\frac{d\sigma}{dt}\right)_{mc} 
\end{equation}
where $Y$ is the yield over $W$ and $Q^2$. The term ``data'' 
refers to the measured experimental yield, and the term ``mc'' 
refers to the simulated events and yield.

To extract the nuclear transparency, which is
defined as the ratio of cross sections extracted 
from data and from a model of pion electroproduction from the 
nucleus without $\pi-N$ final state interactions, simulations were performed
as described in the next section. 

\subsection{Monte Carlo Simulation}
\label{sec:MC}

\begin{figure} 
\begin{center}
\includegraphics[width=75mm]{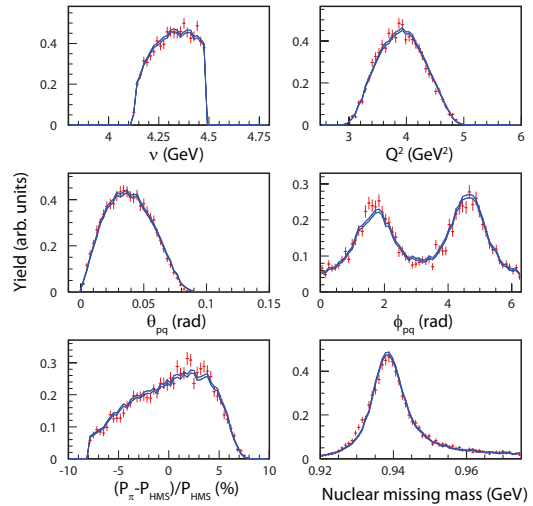}
\caption{\it (color online) Experimental (crosses) and Monte Carlo (lines) distributions for the hydrogen target at $Q^2$=3.91 GeV $^2$.}
\label{fig:8alh2}
\end{center}
\end{figure}

The standard Monte Carlo simulation code for Hall C, SIMC, was
used to simulate the experiment. Events were generated in a phase 
space marginally larger than the acceptance of the spectrometers. 
After events were generated at the vertex, they were transported through 
the spectrometer optics using COSY matrix elements determined from 
a COSY INFINITY~\cite{cosy} model of the spectrometers. Each event was 
weighted by the relevant model cross section (see 
section~\ref{sec:xsecmodel}). A comparison  between experimental and 
simulated distributions of reconstructed quantities for a hydrogen target 
are shown in Fig.~\ref{fig:8alh2}. If the detector setup and the spectrometer 
acceptances (including coincidence acceptance) is realistically modeled in
a simulation, the boundaries of the distributions should match. Differences in 
magnitude can be attributed to differences between the actual cross section 
and the one used in the model. 

To describe electroproduction from nuclear targets the quasi-free approximation
was used, since the energy of the incoming electron is large compared
to the energy associated with the binding of the nucleons. Properties of
the nucleons inside the nucleus are assumed to be described by an
independent particle shell model, where each nucleon interacts with a
mean field exerted by the other nucleons. The probability of finding a
nucleon with momentum ${\bf p}_m$, and separation energy $E_m$, in
the nucleus is given by a spectral function, $S(E_m,{\bf p}_m)$. Existing 
spectral functions were used for deuterium, carbon, gold~\cite{Makins,Neill}, 
and aluminum~\cite{Quint}. The copper spectral function was constructed from 
the iron spectral function described in Ref.~\cite{Makinsth} by increasing the 
number of protons in the outermost 1$f$ shell from 6 to 9 and changing the 
central binding energy of this shell using the separation energy for copper. The
 spectral functions did not include any corrections to account for shifts in 
their strength to large missing momentum caused by nucleon-nucleon correlations.

\begin{figure} 
\begin{center}
\includegraphics[width=75mm]{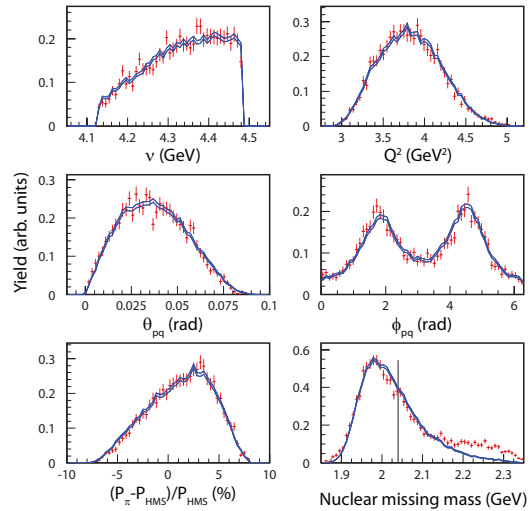}
\caption[Deuterium distributions at $Q^2$=3.9 GeV$^2$.]
{\it (color online) Experimental (crosses) and Monte Carlo (lines) distributions for the deuterium target at $Q^2$=3.91 GeV$^2$. The vertical line in the bottom right panel shows the position of the two-pion production missing mass cut. }
\label{fig:8ald2}
\end{center}
\end{figure}

\begin{figure} 
\begin{center}
\includegraphics[width=75mm]{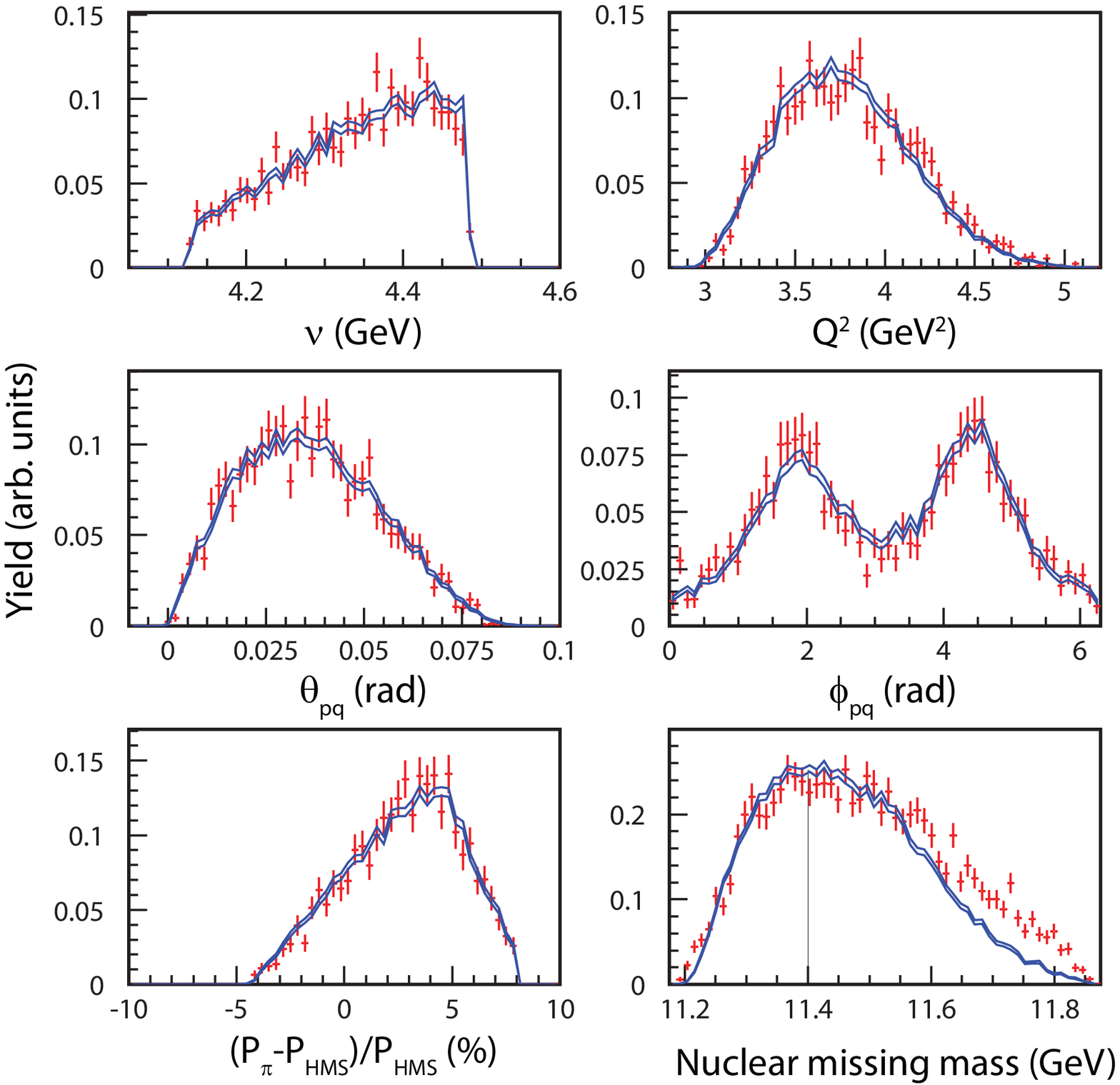}
\caption[Carbon distributions at $Q^2$=3.91 GeV$^2$.]
{\it (color online) Experimental (crosses) and Monte Carlo (lines) distributions for the carbon target at $Q^2$=3.91 GeV$^2$. The vertical line in the bottom right panel shows the position of the two-pion production missing mass cut (see~\ref{sec:multipion} for discussion).}
\label{fig:8acarbon}
\end{center}
\end{figure}

\begin{figure} 
\begin{center}
\includegraphics[width=75mm]{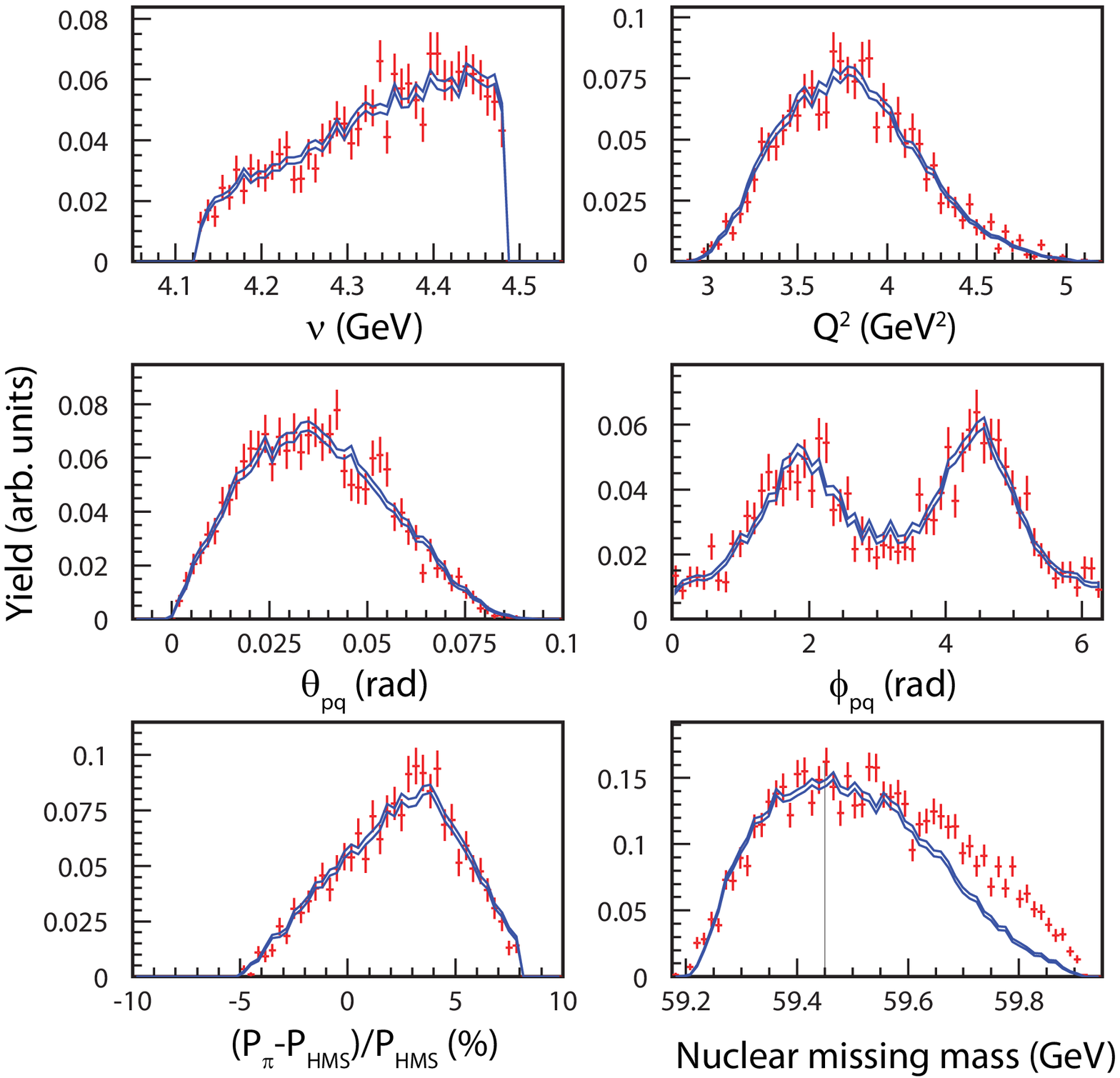}
\caption[Copper distributions at $Q^2$=3.91 GeV$^2$.]
{\it (color online) Experimental (crosses) and Monte Carlo (lines) distributions for the
copper target at $Q^2$=3.91 GeV$^2$. The vertical line in the bottom right panel shows the
position of the two-pion production missing mass cut (see~\ref{sec:multipion} for discussion).}
\label{fig:8acopper}
\end{center}
\end{figure}

\begin{figure} 
\begin{center}
\includegraphics[width=75mm]{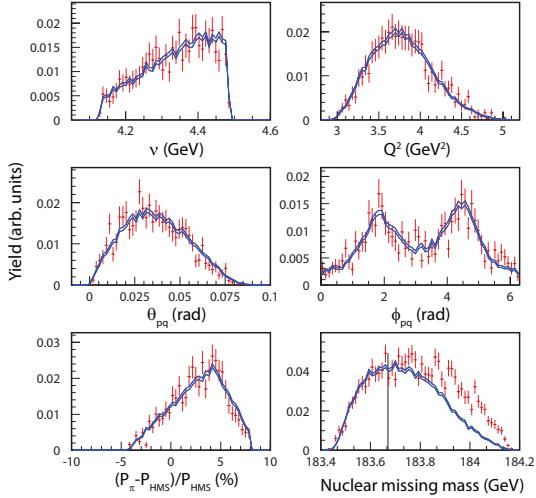}
\caption[Gold distributions at $Q^2$=3.9 GeV$^2$.]
{\it (color online) Experimental (crosses) and Monte Carlo (lines) distributions for the gold target at $Q^2$=3.91 GeV$^2$. The vertical line in the bottom right panel shows the position of the two-pion production missing mass cut (see~\ref{sec:multipion} for discussion).}
\label{fig:8agold}
\end{center}
\end{figure}

Although the momentum of the proton is given by ${\bf p}_m$, $E_m$ is 
not constrained by any of the assumptions in the quasi-free approximation. 
In the present analysis, the off-shellness was described by,
\begin{eqnarray}
  M_A & = &  M_p +M^*_{A-1} \nonumber \\ 
  E_p & = &  M_A -\sqrt{(M^*_{A-1})^2+|{\bf p}_m|^2}
\end{eqnarray}
where $M_A$ is the nuclear missing mass, $M_p$ is the mass of the nucleon, 
$M^*_{A-1}$ is the mass of the spectator nucleons, and $E_p$ is the energy of 
the struck proton. 
A comparison between experimental and simulated distributions of 
reconstructed quantities for nuclear targets is shown in in 
Figures~\ref{fig:8ald2}, ~\ref{fig:8acarbon}, ~\ref{fig:8acopper}, and
~\ref{fig:8agold}.

The radiative corrections used in this analysis are based on the 
formalism of Ref.~\cite{Mo}, which was explicitly modified and improved for ($e,e^\prime p$) 
coincidence reactions in Ref.~\cite{ent}. It includes both external
and internal radiation. The formalism was further modified for pion 
electroproduction, where the target particle is assumed to be a stationary 
proton, while the final pion is assumed to be off-shell. 
The assumption of off-shellness was tested in the simulation and was found to change the 
Monte Carlo yield by at most 0.5\%. The largest source of uncertainty
in the simulation of radiative processes comes from radiation due to the pion
as the electron radiation is relatively well known. The simulated yield changed
by 2-4\% when radiation from the pion was turned off depending on $A$ and 
$Q^2$. For the present analysis. we assume a global uncertainty due to 
radiative correction of 2\%. The uncorrelated uncertainty in the radiative 
corrections was estimated from the target dependence of the simulated yield 
when the pion radiation was turned off, and was found to be 1\% at low $Q^2$ 
and 2\% at high $Q^2$.

SIMC incorporates the effects of the pion decay, multiple scattering and 
energy loss. For the present analysis, 2.5\% (1.4\%) of pion decays occur 
inside a magnet at the lowest (highest) $Q^2$ setting. The random uncertainty 
is about 0.5\% and mainly accounts for muons coming from pions normally 
outside the acceptance. The difference between targets is smaller than 0.1\%. 
The Monte Carlo also takes into account pions which punched through the 
spectrometer collimators. The simulation is based on the calculation of 
the pion transmission through materials described in Ref.~\cite{blok08}.

For the present analysis, SIMC was modified to take into account the effect 
of Pauli blocking. This correction which is determined using the distribution 
function calculated by Fantoni and Pandharipande~\cite{Fantomi} is applied as 
a weight to each event. 

The final state interactions between the knocked-out neutron and the residual nucleons (n-N FSI) can effect the quasi-free cross section and can shift strength in the missing mass spectrum near the single pion production threshold. It is likely to have the strongest effect when the relative momentum between the recoiling neutron and the spectator nucleons is small. An earlier experiment on light nuclei at JLab has shown that the effect of n-N FSI on the quasi-free cross-section reduces with increasing $Q^2$~\cite{gaskellth}. Since this experiment was conducted at relatively high $Q^2$ the n-N FSI was not explicitly accounted for in the simulation. One can account for a fraction of the shift in strength in the missing mass spectrum caused by n-N FSI, by setting the missing energy $E_m$ = 0, when calculating the energy of the proton. However, the lack of full accounting of the n-N FSI is most likely to be the cause of the discrepancy in the shape of the missing mass spectrum at the lowest $Q^2$ = 1.1 $GeV^2$  setting, seen in Fig.~\ref{fig:mm_carbon_missmass} and ~\ref{fig:mm_copper_missmass}.

Coulomb corrections to the incoming and scattered electron are applied 
according to the Effective Momentum Approximation (EMA) approach~\cite{Aste}. 
This approach includes an improvement over earlier versions of the EMA, using an average potential to account for the focusing of the incoming electron wave function~\cite{Aste}. No Coulomb corrections were applied to hydrogen and deuterium, since this 
effect is already included in the elementary pion cross section. The 
corrections for the copper target ranges between 0.2-2\%, while the 
yield correction for the gold target ranges between 0.9-4.4\%. 
We assume that 25\% of this correction contributes to the uncorrelated 
uncertainty.

\subsubsection{Detector acceptances in SIMC}
\label{sec:acceptance}

The acceptances for both spectrometers used in this analysis are within 
the previously determined safe boundaries~\cite{blok08}. The
phase space was studied by comparing distributions for reconstructed
quantities like $Q^2$, $t$, and missing mass as shown in 
Figure~\ref{fig:8alh2}.

The acceptance uncertainties were estimated using the over constrained
H($e,e^\prime p$) reaction. These uncertainties arise from 
the uncertainty in the knowledge of the momentum and angle setting of 
the spectrometers and the beam energy. 
The point-to-point yield variation was found to be 1\% for
different $Q^2$ values.

\subsubsection{The model cross section}
\label{sec:xsecmodel}

The Monte Carlo was used to extract the bin-centered
experimental cross section by iterating the model cross section until
Monte Carlo distributions matched the data.
The starting model was based on the previous pion electroproduction
data from Hall C~\cite{blok08}.  The model cross section was
taken as the product of a global function describing the
$W$-dependence times (a sum of) $Q^2$ and $t$ and $\theta$ dependent
functions for the different structure functions. A correction function,
which was assumed to factorize,
\begin{equation}
    C_H(W,Q^2,t,\phi_{pq}) = O(W)K(Q^2)T(t)F(\phi_{pq}),
\end{equation}
was determined by iterating the model and comparing it to the hydrogen 
elementary cross section. These correction functions were assumed to
be second order polynomials, with the exception of $F(\phi_{pq})$, which
was assumed to be of third order.

\subsubsection{Iteration procedure}
\label{sec:iter}

For the hydrogen target the bin centered experimental cross section is 
given by
\begin{equation}
\left(\frac{d^5\sigma_A}{d\Omega_{e'}dE_{e'}d\Omega_{\pi}}\right)^{exp}_{x_0}
= \frac{Y_{data}}{Y_{MC}}
\left(\frac{d^5\sigma_A}{d\Omega_{e'}dE_{e'}d\Omega_{\pi}}\right)^{model}_{x_0}. \end{equation}
For nuclear targets, the bin centered experimental cross section is given by
\begin{equation}
\left(\frac{d^6\sigma_A}{d\Omega_{e'}dE_{e'}d\Omega_{\pi}dP_{\pi}}\right)^{exp}_{x_0}
= \frac{Y_{data}}{Y_{MC}}
\left(\frac{d^6\sigma_A}{d\Omega_{e'}dE_{e'}d\Omega_{\pi}dP_{\pi}}\right)^{model}_{x_0},
\end{equation}
where the subscript, $x_0$, indicates that the cross section is evaluated at 
a particular point $(W_0,Q^2_0,\theta_0)$ inside the acceptance. 
The model cross sections for hydrogen and for the nuclear targets were 
determined using a point Monte Carlo simulation, which was
performed using scattered electron kinematics and pion angles generated 
randomly within a very narrow phase space volume that corresponds 
to $W_0$, $Q^2_0$ and $\theta_0$. For nuclear targets, $P_{\pi}$ is 
generated over the whole phase space from which a narrow range was selected
in the analysis. 

The extracted cross sections depend on the initial cross section model, 
and thus there are systematic uncertainties associated with it. This uncertainty is obtained by extracting the cross section using a different starting
model and was found to be 1.1\%. However, this uncertainty will not contribute to nuclear transparency, since nuclear transparency involves the ratio of two Monte Carlo yields.

\subsubsection{Multi pion production in nuclear targets}
\label{sec:multipion}

The quasi-free pion production model is limited to single pion production. The
production of more than one pion in a single event (multiple pion production)
was suppressed for hydrogen targets during the pionCT experiment due to the 
relatively high $Q^2>$ 1 GeV$^2$ and $W>$2.1 GeV being above the resonance 
region. This suggests that the mechanism for multiple-pion production involves
the outgoing pion producing one or more pions from a nucleon in a second process
that was incoherent from the production of the first pion. Multiple-pion
events can only be produced above a missing mass threshold that is larger than
the missing mass threshold for single-pion production, i.e., 
$M_x=M_{A-1}+M_{\pi}$ for a nucleus of mass $A$. 
Indeed, in the analysis, at first a missing mass cut exactly corresponding to 
this threshold was used to suppress multiple-pion events. However, this cut 
resulted in an unacceptable loss of statistics at the highest $Q^2$ settings, 
and thus an alternative cut on the nuclear missing mass above the multi-pion threshold were used. 
\begin{figure}
\begin{center}
\includegraphics[width=75mm]{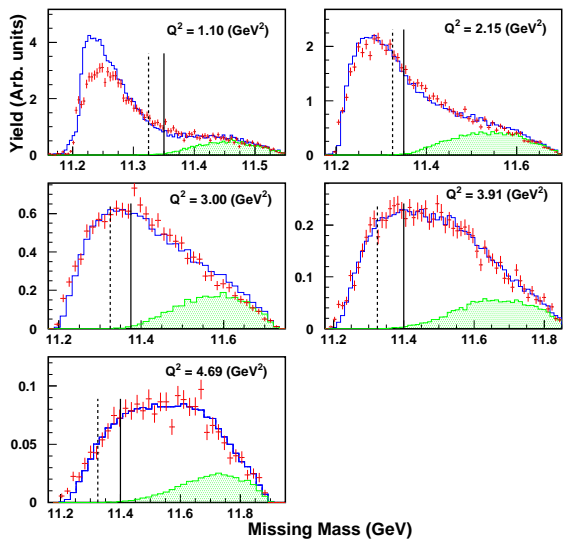}
\caption[Carbon nuclear missing mass with multiple-pion contribution.]
{\it (color online) Nuclear missing mass distributions (in GeV$^2$) for $^{12}$C(e,e$'\pi^+$)
The data (red crosses) are compared to the simulation (blue line) which is a sum of single-pion  and multiple-pion simulations. The shaded area (green) shows the contribution from the multi-pion simulation.  The full simulation is normalized to the data. The dashed vertical lines represents the threshold for double-pion production (11.34 GeV$^2$).  The solid lines represent the position of the cut used in this analysis.}
\label{fig:mm_carbon_missmass}
\end{center}
\end{figure}

In order to describe events above the two pion threshold, a  
multiple-pion production simulation was developed for the present nuclear
target analysis. The mechanism for multiple-pion production was assumed to be
quasi-free single-pion production from a nucleon followed by a secondary 
process that was incoherent from the first, where the pion produces one or 
more pions from a different nucleon. The cross section for the second process 
was assumed to be uniform over the acceptance of the HMS spectrometer. 
\begin{figure}
\begin{center}
\includegraphics[width=75mm]{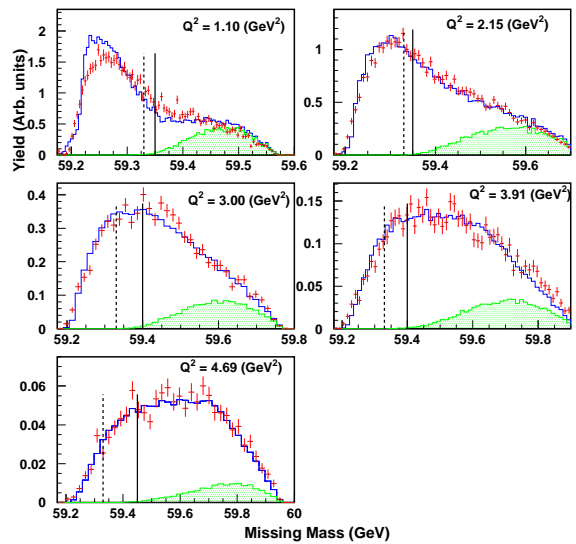}
\caption[Copper nuclear missing mass with multiple-pion contribution.]
{\it (color online) Nuclear missing mass distributions (in GeV$^2$) for $^{63}$Cu(e,e$'\pi^+$). The data (red crosses) are compared to the simulation (blue line) which is a sum of single-pion  and multiple-pion simulations. The shaded area (green) shows the contribution from the multi-pion simulation.  The full simulation is normalized to the data. The dashed vertical lines represents the threshold for double-pion production (59.33 GeV$^2$).  The solid lines represent the position of the cut used in this analysis. }
\label{fig:mm_copper_missmass}
\end{center}
\end{figure}

The effect of multi-pion production can be seen in 
Fig.~\ref{fig:mm_carbon_missmass} and ~\ref{fig:mm_copper_missmass} for the 
carbon and copper targets. The agreement between the missing mass distributions obtained from data and simulation improves with increasing $Q^2$. The discrepancy seen at $Q^2$ = 1.1 $GeV^2$ is attributed to the reaction mechanisms missing from the simulation such as final state interactions between the knocked-out neutron and the residual nucleons (nN-FSI) and short range correlations. The effect of these reaction mechanisms reduces with increasing $Q^2$. 

These results show that it is safe to increase the double-pion missing mass cut above the threshold with minimal contamination. The double-pion missing mass cut was placed at the position where the systematic uncertainty from the contribution of multiple-pion events was less than 5\%. With these cuts, the total uncertainty due to multi-pion contamination is $<$ 0.4\%. We also noted an interesting  smooth $A$ dependence in the ratio of the multiple-pion to single-pion yields.

\subsubsection{Test of the quasi-free assumption}
\label{sec:ltsep} 

\begin{figure} 
\begin{center}
\includegraphics[width=75mm]{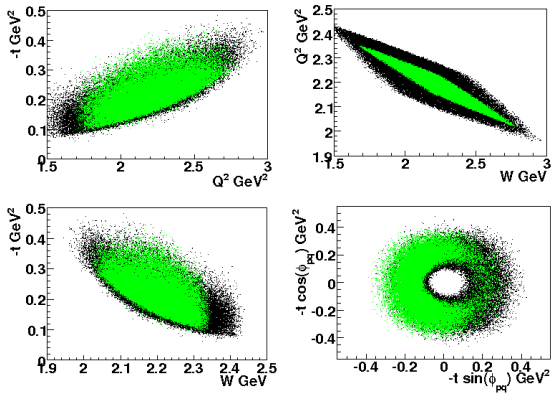}
\caption[Phase space for high and low $\epsilon$ data:]
{\it (color online) The plot on the upper left panel is the
  phase space comparison between the high (black) and low (blue)
  $\epsilon$ for $-t$ 
  vs $Q^2$. The plot on the upper right panel is the phase
  space comparison for $W$ vs $Q^2$. The plot on the lower left
  panel shows the comparison  for $-t$ vs $W$. The lower right panel shows the
  azimuthal coverage for these two data sets.} 
\label{fig:phase}
\end{center}
\end{figure}

\begin{figure} 
\begin{center}
\includegraphics[width=75mm]{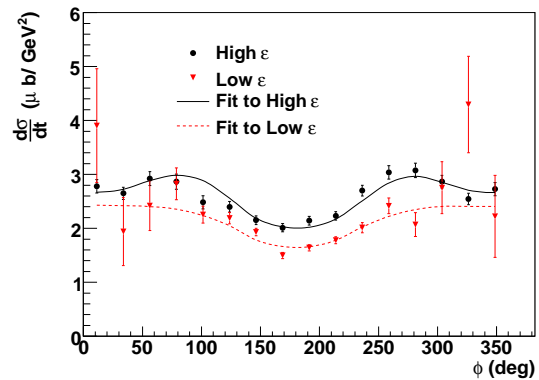}
\caption[Phase space for high and low $\epsilon$ data:]
{\it (color online) Representative plot of the experimental cross sections,
$\frac{d^2 \sigma}{dt d\phi}$ as a function of the azimuthal angle
  $\phi$ at $Q^2$=2.15 GeV$^2$ for high and low $\epsilon$. The curves
  shown represent the fit of the measured values of the cross section
  to equation~\ref{eq:sepsig1}. The $-t$ in this plot corresponds to the common region between the high and low $\epsilon$ data shown in Fig.~\ref{fig:phase}. Only the statistical uncertainties are shown.}
\label{fig:phi}
\end{center}
\end{figure}

The average cross sections were extracted by
integrating over the whole acceptance ($W$, $Q^2$  and $t$). This
averaging reduces the systematic uncertainties related to the cross
section model in the Monte Carlo by smearing the exact kinematic information of
the extracted cross section. We bin the data in $\phi$ by
integrating over all other kinematic variables at each $\epsilon$
setting. The azimuthal angular coverage for the hydrogen target at $Q^2$ =
2.15 GeV$^2$ can be found in Fig.~\ref{fig:phase}. Due
to the correlation in the kinematics, the central $W$, $Q^2$ and $t$
values for each $\phi$ bin are different from those obtained by integrating 
over the entire $\phi$ region. Thus, a Monte Carlo simulation with model
cross sections is used to account for the correction between the
cross section evaluates at the center of each $\phi$ bin and the one for the 
entire $\phi$ region. 
The four structure functions in equation~\ref{eq:sepsig1} are extracted by 
fitting the data with respect to $\phi$ for both high and low
$\epsilon$ settings simultaneously.
A representative fit for the hydrogen target at $Q^2$ = 2.15 GeV$^2$ can be 
found in Fig.~\ref{fig:phi}. In the analysis, an additional acceptance 
cut is used to ensure that the kinematic region given by $W$, $Q^2$ and $t$
is the same at high and low $\epsilon$. Such a phase space
comparison is shown in Fig.~\ref{fig:phase}.

\begin{figure}
\begin{center}
\includegraphics[width=80mm,height=55mm]{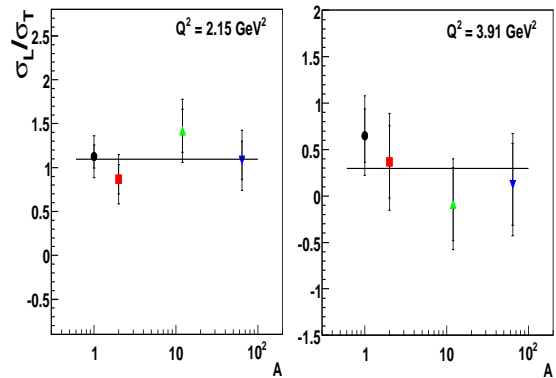}
\caption{\it (color online) The ratio of longitudinal to transverse
  cross sections for pion electroproduction from $^1$H, $^2$H,
  $^{12}$C, and $^{63}$Cu targets, at fixed $Q^2$ = 2.15 (left) and 3.91 (right)
  GeV$^2$. The inner error bars represent the statistical 
  uncertainties, while the outer error bars are the sum in
  quadrature of the statistical and systematic uncertainties. The curves 
  represent constant-value fits to all nuclear data at a fixed $Q^2$. The probability of these constant-value fits assuming Gaussian statistics is 69 and 70\%, respectively.} 
\label{fig:qs}
\end{center}
\end{figure}

\begin{tiny}
\begin{table*}
  \renewcommand{\arraystretch}{1.2}
  \centering
    \begin{tabular}{||l|c|c|c||}
      \hline
      Item & Uncorrelated uncertainty& $A$ dependent &Correlated uncertainty\\
      & ($\%$) & ($\%$) & ($\%$)\\
      \hline
      \hline
      HMS \v{C}erenkov     & 0.2    &  & 0.3-0.5      \\
      SOS \v{C}erenkov     & 0.2    &  & 0.3-0.5     \\
      Charge               & 0.4-0.9&  & 0.4          \\
      Coincidence blocking & 0.2    &  &             \\
      HMS Trigger          & 0.5    &  &             \\
      Dead time            & 0.2-0.5&  &             \\
      HMS Tracking         & 1.0    &  & 1.0         \\
      SOS Trigger          & 0.5    &  &             \\
      SOS Tracking         & 0.5    &  & 0.5         \\
      Pion decay           & 0.1    &  & 1.0         \\
      Coulomb corrections  & $<$1.0 &  &             \\
      Radiative corrections &1.0-2.0&  & 2.0       \\
      Collimator           & 0.5    &  & 1.0         \\
      Acceptance           & 1.0    &  & 2.0         \\
      Iteration procedure  & 1.1    &  &            \\
      Multi-pion contamination &$<$0.4& &             \\
      Target thickness     &   & 0.5 - 1.0&       \\
      Pion absorption      &   & 0.5    & 2.0         \\ \hline
      Total                & 2.4-3.4 &0.7-1.1 & 3.9-4.0     \\
      \hline
      model dependence     & 3.5-7.6 &    &    \\
      \hline
    \end{tabular}
  \caption[Systematic uncertainties]
       {\it The systematic uncertainties in extracting nuclear transparency. The uncorrelated uncertainty contributes directly to the point-to-point uncertainty in the nuclear transparency, the $A$ dependent uncertainty is independent of $Q^2$, while the correlated uncertainty is independent of both the target nucleus and $Q^2$. }  
 \label{table:sys}
\end{table*}
\end{tiny}

\begin{tiny}
\begin{table*}
  \renewcommand{\arraystretch}{1.2}
  \centering
    \begin{tabular}{||l|c|c||}
      \hline
      Item  & Uncertainty in &
      Uncertainty in \\
            & Differential cross section &  L-T cross section ratio \\
      & ($\%$) & ($\%$)\\
      \hline
      \hline
      HMS \v{C}erenkov        & 0.4-0.5 & \\
      SOS \v{C}erenkov        & 0.4-0.5 & \\
      Charge                  & 0.6-1.0 & \\
      Coincidence blocking    & 0.2     &\\
      HMS Trigger             & 0.3-0.5 &\\
      Dead time               & 0.2-0.5 &\\
      HMS Tracking            & 1.1-1.4 &\\
      SOS Trigger             &  0.3    &\\
      SOS Tracking            & 0.5     &\\
      Pion decay              & 1.0     &\\
      Coulomb corrections     & $<$1.0  &\\
      Radiative corrections   & 2.2-2.8 &\\
      Collimator              & 1.1     &\\
      Acceptance              & 2.2     &\\
      Iteration procedure     & 1.3 - 1.5    & 13.0-18.0\\
      Multi-pion contamination & $<$0.4  &\\
      Target thickness        & 0.5-1.0 &\\
      Pion absorption         & 2.1     &\\
      Kinematics               & 1.5-2.0 & 3.0-16.6 \\
      momentum coverage        &         &  $<$12.0\\
      \hline
      Total                    & 4.8-5.7 & \\
      \hline
    \end{tabular}
  \caption[Systematic uncertainties]
       {\it The systematic uncertainties in extracting cross sections
       and in extracting the ratio of longitudinal to transverse 
cross section}  
 \label{table:sys1}
\end{table*}
\end{tiny}
The same fitting procedure described above could be used to
obtain the Rosenbluth-separated pion electroproduction cross sections.
However, where this separation is relatively straightforward for hydrogen
targets, a similar separation for nuclear targets relies on the assumed 
quasi-free reaction mechanism. This is because beyond the $W$, $Q^2$ and $t$ dependence, the elementary off-shell pion electroproduction cross section has 
a $P_{\pi}$ dependence in the nuclear medium. Hence the input model for elementary pion electroproduction needs to starting value for the $P_{\pi}$ dependence. This quasi-free $P_\pi$-dependence is taken into account in the Monte Carlo simulations for the nuclear targets, and the iterative procedure described earlier is followed. This implies that the extracted nuclear cross sections represent the averaged values integrated over a wide kinematic acceptance, rather than the bin centered value. These averaged cross sections are used to obtain the longitudinal to transverse ratio between nuclear and hydrogen targets. The ratio of longitudinal to transverse cross sections at fixed $Q^2$ = 2.15 and 
3.91 GeV$^2$ for the various targets used in this experiment are shown in 
Fig.~\ref{fig:qs}. 
We find no difference, within the experimental uncertainties, between the ratio of longitudinal to transverse cross sections for nuclear and hydrogen targets, this can be viewed as a confirmation of the quasi-free reaction reaction mechanism.

\section{Uncertainty estimates}
\label{Sec:sys}

Table~\ref{table:sys} lists the systematic uncertainties in extracting the
nuclear transparency. Several sources of these uncertainties have
already been discussed in sections~\ref{sec:acceptance}-\ref{sec:multipion}. 

The uncertainty in the acceptance is based on extensive single-arm elastic
and deep-inelastic measurements from~\cite{chr04,tva04}, and 
${^1}$H($e,e^\prime p$) data, 
including sieve-slit data on a carbon target, taken to check the optical matrix elements.
The influence of the uncertainties in the offsets in the kinematical variables
such as beam energy, momenta and angles, were determined by changing their 
values by their uncertainty and evaluating the resultant changes in the 
cross sections.
\begin{figure*}
\begin{center}
\includegraphics[width=160mm,height=110mm]{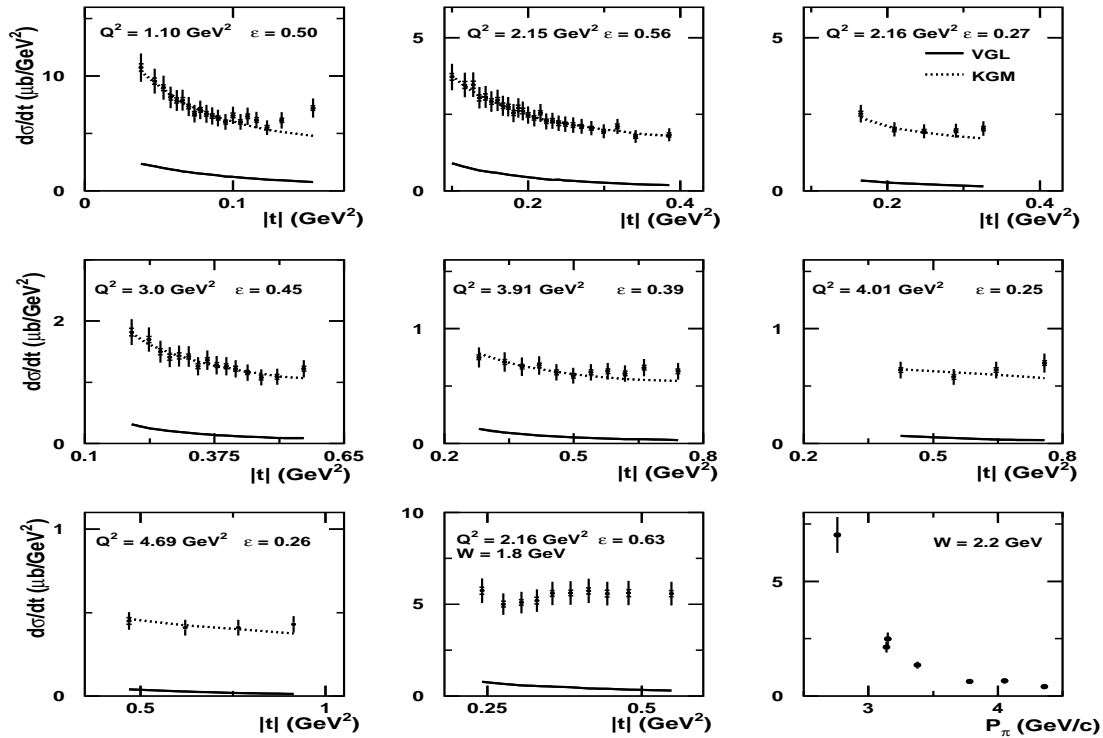}
\caption
{\it Differential cross sections $\frac{d\sigma}{dt}$ for pion
  electro-production from hydrogen versus $|t|$. The last panel (bottom right) 
shows the differential cross section versus the pion lab momentum (only the seven $W=$2.2 GeV points are shown here). For each of the points shown in the 
cross section versus pion momentum plot (last panel) the data were averaged over the respective $t$ range shown in the previous panels. For the panels showing differential cross section versus $|t|$ the center of
  mass energy is $W$ = 2.2 GeV for all except one kinematic setting where
  $W$ = 1.8 GeV (bottom middle). The inner error bars represent the statistical
  uncertainties, while the outer error bars are the sum in quadrature
  of the statistical and systematic uncertainties. The data are
  compared with both the VGL-Regge~\cite{VGL} and the KGM~\cite{Mosel}
  calculations where available (see text).} 
\label{fig:hdxs}
\end{center}
\end{figure*}
\begin{figure*}
\begin{center}
\includegraphics[width=160mm]{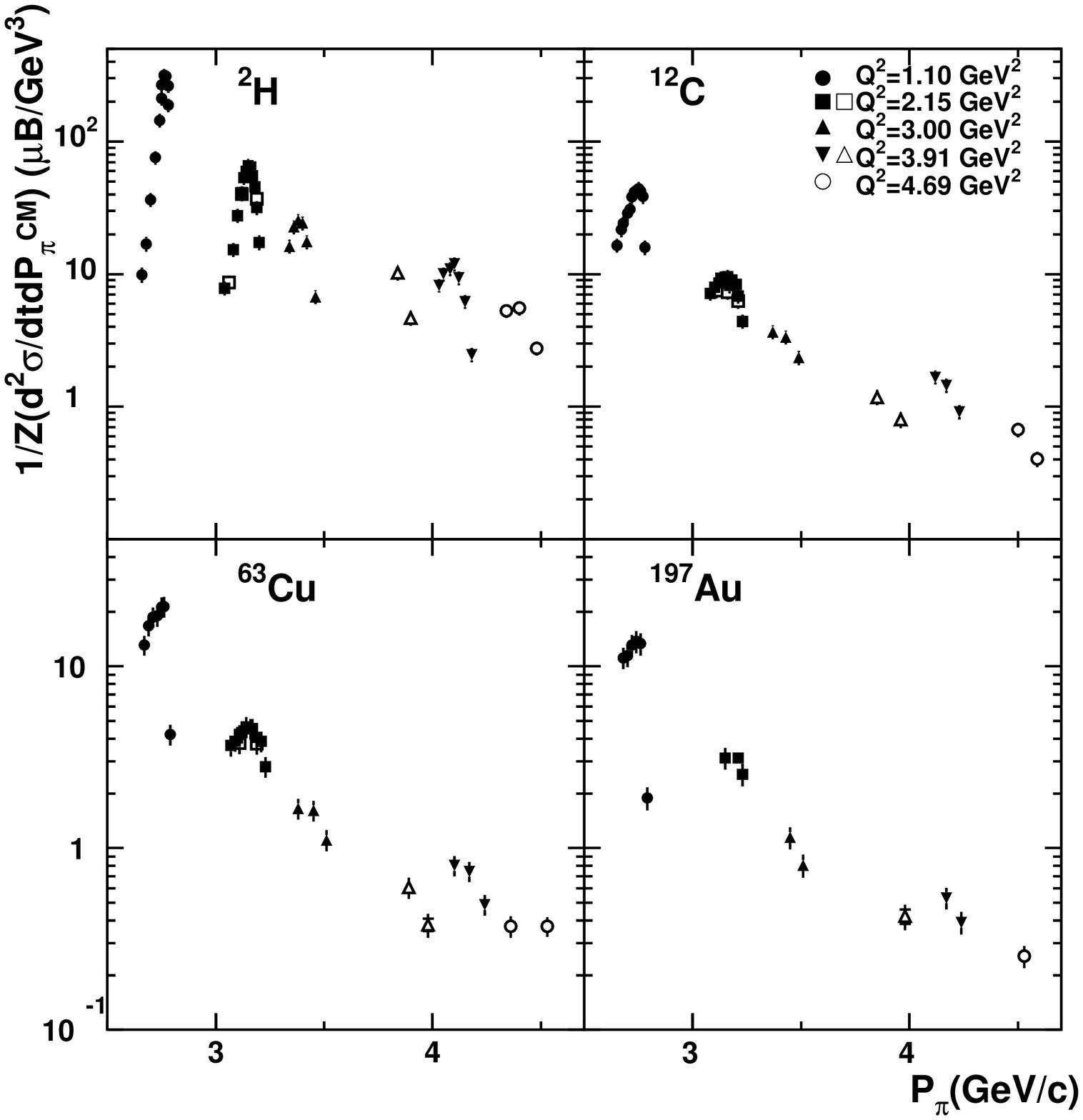}
\caption
{\it The extracted doubly-differential cross sections
  $\frac{1}{Z}\frac{d\sigma}{dtdP^{CM}_{\pi}}$ versus the pion
  momentum $P_{\pi}$ for the deuterium, carbon, copper, and gold
  targets, respectively. The cross sections are
  normalized by Z, since $\pi^+$ can only be generated from a proton in
  this channel. The inner error bars represent the
  statistical uncertainties, while the outer error bars are the sum in
  quadrature of the statistical and systematic uncertainties. For each
  target at Q$^2$ =2.15 GeV$^2$ and Q$^2$ = 3.91 GeV$^2$  the solid
  symbol represents the high $\epsilon$ kinematics while the open
  symbols represent the low $\epsilon$ kinematics.}   
\label{fig:target}
\end{center}
\end{figure*}  

The uncertainty in the target thickness for the solid targets is
dominated by the knowledge of their purities and the thicknesses. For the 
4 cm liquid targets, the uncertainty is dominated by the target boiling 
correction and the beam path length since the beam did not pass through the 
axis of the liquid target cylinder. The uncertainty in the total charge 
includes a 0.4\% uncertainty due to calibration of the beam 
current monitors. 

The largest correlated systematic uncertainties are radiative corrections, and acceptance, resulting in a total correlated 
uncertainty of 3.9-4.0\%. 
The uncorrelated systematic uncertainty is dominated by contributions from the acceptance, radiative corrections and the iteration procedure, resulting in a total uncorrelated uncertainty of 2.4 - 3.4\%.
The largest contribution to the ``$A$ dependent'' uncertainty is the 
target thickness, resulting in a total uncertainty of 0.7 to 1.1\%.

The largest source of uncertainty in the radiative correction
procedure comes from pion radiation as the electron radiation is relatively well
known. The Monte Carlo equivalent yields changed by 2-4\% when pion radiation was turned off (2\% at low $Q^2$ and 4\% for the heavy targets at high $Q^2$). From this the normalization uncertainty was taken to be 2\%. The point-to-point uncertainty in the radiative corrections was estimated from the target dependence of the Monte Carlo equivalent yield when the pion radiation was turned off. This was 1\% at the low $Q^2$ and 2\% at high $Q^2$.

In addition to these uncertainties, a $Q^2$ dependent model uncertainty was also determined to be 3.5\% -7.6\%. This uncertainty is the quadrature sum of the change in $Q^2$ dependence of the transparency when using two different spectral functions and two different Fermi distributions in the simulation, and the $Q^2$ dependent uncertainty from reactions mechanisms not included in the simulation. 
The uncertainty from the reaction mechanisms not included in the simulation were estimated by quantifying the difference in shape of the missing-mass spectra 
from data and simulation.

Table~\ref{table:sys1} lists the systematic uncertainties in extracting the unseparated cross sections and the uncertainty in extracting the ratio of longitudinal to transverse cross section. These were obtained using the quadrature sum of the correlated, uncorrelated and $A$ dependent uncertainties along with two additional sources of uncertainty labeled as ``Kinematics'' and ``momentum coverage.''
The ``Kinematics'' part represents the uncertainties in the knowledge of spectrometer angles and momentum setting. It will not 
contribute to the nuclear transparency, since the spectrometer settings are exactly the same for different targets. The ``momentum coverage'' part
represents the effect of the wider momentum cut (in order to obtain
enough statistics in the low $\epsilon$ region) used in the Rosenbluth
separation than the extraction of nuclear transparency.

\section{Results and Discussion}
\label{sec:results}

The differential cross sections for hydrogen are shown in Fig.~\ref{fig:hdxs}
and those for all four nuclear targets (deuterium, 
carbon, copper and gold) are shown in Fig.~\ref{fig:target}. All numerical
values are tabulated in Appendix A and B. In the following sections, the global 
dependencies of the hydrogen and nuclear cross sections will be reviewed and 
the data compared to recent model calculations. The results of the quasi-free reaction mechanism test and the extracted nuclear transparencies are also shown.

\subsection{Global dependencies and model comparison for hydrogen}
\label{sec:dsighyd}

For the different values of $Q^2$, the differential cross section shows the
characteristic fall-off with $-t$, which may be due to the pion pole 
in $\sigma_L$. The magnitude of the cross section decreases 
at constant $W$ and with increasing $Q^2$, mostly because the value
of $-t_{min}$ increases with $Q^2$. In Fig.~\ref{fig:hdxs} (bottom right panel) we show the differential cross section vs pion lab momentum ($P_{\pi}$). In this plot each point at a particular $P_{\pi}$, represents the differential cross section averaged over the $|t|$ range shown in the one of the other panels of Fig.~\ref{fig:hdxs} that corresponds to the $Q^2$ values at that $P_{\pi}$ (see Table ~\ref{table:kine}).

The cross sections are compared to predictions of two different
models of pion electroproduction, the VGL-Regge model~\cite{VGL}, and
the more recent ``KGM'' model~\cite{Mosel}. 
The VGL-Regge calculations are in a gauge invariant model incorporating
$\pi$ and $\rho$ Regge trajectory exchanges. They significantly underestimate
the measured differential cross sections. Most of the discrepancy can likely
be attributed to the model underestimating the transverse part of the 
cross section as shown in Ref.~\cite{Hornt}, while they agree well with 
$\sigma_L$. The recent KGM model~\cite{Mosel}, which includes
a deep-inelastic scattering (DIS) Ansatz for the transverse part of the
cross section, with the longitudinal cross section dominated by hadronic
degrees of freedom and the pion electromagnetic form factor, agrees much
better with the measured differential cross sections.

\subsection{Global dependencies for nuclear targets} 
\label{sec:dsignuc}

We have extracted the differential cross sections for all four nuclear
targets (deuterium, carbon, copper and gold) at the eight different kinematics
settings given in Table I. Here, the additional complication
introduced due to the added degree of freedom induced by the Fermi-motion 
(or more general, the nuclear binding) of the struck proton is taken into 
account by extracting the doubly-differential cross sections
$\frac{d^{2}\sigma}{dtdP^{CM}_{\pi}}$, where $P^{CM}_{\pi}$ is the pion momentum
in the center-of-mass frame of the virtual photon and the nucleus.

The local variations in the nuclear cross sections as illustrated in 
Fig.~\ref{fig:target}
indicate effects due to Fermi motion beyond the central kinematics. These
local variations are more pronounced for the deuterium target, because of
its narrower Fermi cone. 
Although the general trend of the nuclear cross section is similar to that of the hydrogen cross section, the fall-off of the nuclear cross sections
with $P_\pi$ is steeper than that found for the hydrogen cross sections.

\subsection{Verification of the quasi-free mechanism}
\label{sec:quasifree}

A prerequisite for an interpretation of nuclear transparency as a function of $Q^2$ is that the reaction mechanism remains identical over the $Q^2$ range.
This translates into an important condition in searching for CT using
pion electroproduction: the reaction should proceed through a quasi-free mechanism. 

There are many mechanisms that could break down the quasi-free assumption,
such as:

\begin{itemize}
\item{\bf Nucleon-nucleon ($NN$) final-state interactions} 
  The amplitude of $NN$ final-state interactions can interfere with the
  elementary electro-pion production amplitude, which can change the
  ratio of the longitudinal and the transverse cross section. 
  For example, the disagreement between the missing mass spectra from data and simulation at $Q^2$ = 1.1 (GeV)$^2$ (Figs.\ref{fig:mm_carbon_missmass} and ~\ref{fig:mm_copper_missmass}) is expected to be due to such effects, and other
  potential reaction mechanism effects not included in the Monte Carlo model.
  The agreement between data and Monte Carlo at higher values of $Q^2$,
  however, suggests only a small contribution from such $NN$ final-state
  interaction complications.
\item{\bf Rescattering}
  Rescattering involves the electro-production of a meson followed by a
  second interaction that produces the detected $\pi^+$ particle. For example,
  such rescattering contributions have been shown to dominate the cross section
  in $\rho^0$ photo-production for $t \geq$ 0.5 (GeV)$^2$~\cite{clasrho}. 
  In principle, if the rescattering effect dominates, one would expect a
  modification of both the longitudinal to the transverse cross section ratio,
  and the cross section dependence on $W$. 
\item{\bf Pion excess} 
  Excess pions may be present in a nuclear system due to the long range of
  meson-exchange currents~\cite{Friman}. If such pion excess effects are
  significant, one would anticipate a change in the ratio $\sigma_L/\sigma_T$
  of the measured nuclear cross sections as compered to the hydrogen
  cross sections. An earlier experiment on light nuclei at low $Q^2$ did not
find any pion excess~\cite{gaskell}.
\item{\bf Medium modification of nucleons}
  The European-Muon Collaboration (EMC) discovered \cite{EMC} that the
  structure functions for deep inelastic inclusive lepton scattering off
  nuclear targets differed from deuteron targets. Although the effect remains
  poorly understood, it is generally accepted that nucleon structure will be
  modified within a nuclear medium, both due to nuclear binding and to
  non-nucleonic QCD effects. Such a medium modification of nucleons could
  also impact the nuclear pion electro-production cross sections, but is
  generally expected to also lead to a change in the ratio of longitudinal
  and transverse pion electro-production cross sections, $\sigma_L/\sigma_T$.
\item{\bf Two-nucleon correlations}
  A series of A($e,e^\prime p$) measurements revealed that spectroscopic
  factors for proton valence shells were quenched by approximately 30-35$\%$
  as compared to mean-field expectations~\cite{Frois,Lapikas}. A possible
  explanation for this discrepancy is that correlations move some of the
  single-particle strength to orbitals above the Fermi energy. This kind of
  correlation will change the nucleus spectral function, and thus break down
  the quasi-free assumption. 
\end{itemize} 

The most straightforward verification of the quasi-free mechanism 
is the equivalence of the longitudinal-transverse character for pion 
electroproduction from nuclear and hydrogen targets. 
The ratio of longitudinal to transverse cross sections should, for instance, 
be independent of the nuclear atomic number $A$. 
The ratio of longitudinal to transverse cross sections at fixed $Q^2$ = 2.15 and 
3.91 GeV$^2$ for the various targets in this experiment are shown in 
Fig.~\ref{fig:qs}. 

The $A$-dependence of the $\sigma_L/\sigma_T$ ratio 
agrees with a flat line fit within the experimental uncertainties, and are 
thus consistent with the quasi-free assumption. However, 
they cannot rule out non quasi-free reaction mechanism effects that affect
the longitudinal and transverse character of pion electroproduction in a
similar fashion, 
for instance,  $NN$ final-state interaction, rescattering, pion excess, or
medium modification effects. Note that we have intentionally kept the value of
$-t$ of the measurements low ($\le$ 0.5 GeV) to minimize complications due
to rescattering or two-nucleon effects. Together with the overall good
agreement of data and Monte Carlo simulations (beyond $Q^2$ = 1.1 GeV$^2$), 
we have gained confidence in the validity of the quasi-free reaction
mechanism.

\subsection{Nuclear Transparencies}
\label{sec:transp}

As mentioned earlier, nuclear transparency is defined as the ratio of the cross section per nucleon for a process on a bound nucleon inside a nucleus to that from a free nucleon. The $P_{\pi}$ dependence of the nuclear transparencies is shown in 
Fig.~\ref{fig:nt}. Results are shown for deuterium and carbon (top left),
copper (bottom left) and gold (bottom right) and aluminum (top right), which 
was also used for determining the cell wall background contribution for 
cryogenic targets. 

Before we discuss the various model calculations, we first redefine
the nuclear transparency ($T_D$) as the cross sections of heavy nuclear targets
as compared to those of a deuterium target. This reduces the uncertainty due to both the unknown elementary pionelectroproduction off a neutron and effects due to the procedure
to take into account Fermi motion. The results are shown in
Fig.~\ref{fig:ntld2}, where we present the data versus $Q^2$ rather
than pion momentum P$_\pi$. Of course, the results are not too different,
as could be anticipated from the nuclear transparency results of deuterium
to be close to unity, as can be seen in top left panel of Fig.~\ref{fig:nt}.
We note in addition that the deuteron nuclear transparency results are
found to be consistent with a constant-line fit with 81\% probability.
The nuclear transparency results are listed in Appendix C.

The nuclear transparencies are expected to be near-constants over the
pion momentum range of the experiment from traditional nuclear physics
point of view~\cite{glauber}, because the hadron-nucleon cross sections are
nearly independent of momentum over the range of momenta in Figs.~\ref{fig:nt} and~\ref{fig:ntld2}. Instead, the observed pion nuclear
transparency results (as compared both to hydrogen and deuterium
cross sections) shows a slow but steady rise versus pion momentum for
the nuclear ($A >$ 2) targets, causing a deviation from calculations 
without CT. 

\subsection{Comparison with model calculations}
\begin{figure}
\begin{center}
\includegraphics[width=90mm,height=85mm]{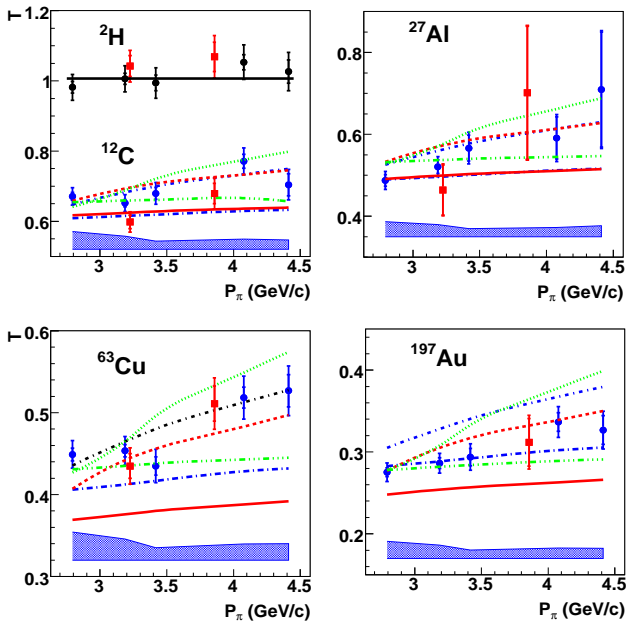}
\caption
{ \it (color online) Nuclear transparency, T, vs $P_{\pi}$ for $^2$H and $^{12}$C, $^{27}$Al, $^{63}$Cu, and $^{197}$Au. The inner error bars are the statistical
  uncertainties and the outer error bars are the statistical and
  point-to-point systematic uncertainties added in quadrature. The solid circles (blue) are the high $\epsilon$ points while the solid squares (red) are low $\epsilon$ points. The dark (blue) bands are the model uncertainties 
(for details see  Sec.~\ref{Sec:sys}). The dashed and solid lines (red) are 
Glauber calculations  from Larson, et al.~\cite{Larson06ge}, with and without CT respectively. Similarly, the dot-short dash and dot-long dash lines (blue) are Glauber calculations with
  and without CT from Cosyn, et al.~\cite{Cosyn}. The effects of short-range correlations are included in these latter calculations. The dotted and dot-dot-dashed lines (green) are microscopic+ BUU transport calculations from Kaskulov et al.~\cite{mosel2}, with and without CT respectively.}  
\label{fig:nt}
\end{center}
\end{figure}

We compare our results with the calculations of Larson, {\it et al.}~\cite{Larson06ge} (solid and dashed curves), Cosyn, {\it et al.}~\cite{Cosyn} (dot-short dash and dot-long dash curves) and Kaskulov, {\it et al.}~\cite{mosel2}(dotted and dot-dot-dash).

Larson, {\it et al.} compute the nuclear transparency at the exact 
kinematics of the experiment in terms of semi-classical formula based on the 
eikonal approximation and a parametrization of the effects of 
final state interactions (FSI) in terms of 
an effective interaction. This semi-classical formula involves a single 
integral over the path of the outgoing pion which is suited for situations in 
which the kinematics of the final pion are known. The nuclear density is taken 
as a Woods-Saxon form with radius parameter $R= 1.1 A^{1/3}$~fm and diffuseness
 $a=0.54$~fm. The effective interaction is based on the quantum diffusion model
 of Ref.~\cite{Farrar}, which predicts the interaction of the PLC to be 
approximately proportional to the propagation distance $z$ for $z < l_c$
The coherence length (or formation length) is parametrized as described earlier
in section~\ref{sec:transpmethods}. Larson, Miller and Strikman 
use the following parameters; $T_{lifetime}$ = 1 fm/c 
and $M_h^2$ = 0.7 (GeV/c$^2$)$^2$. In the limit of $l_c = 0$ a PLC is not 
created and the effective interaction reduces to a Glauber-type calculation 
with $\sigma_{eff} \approx \sigma_{\pi N}(P_{\pi})$, the $\pi -N$ cross section 
for pion momentum $P_{\pi}$ obtained from
a parametrization by the Particle Data Group (PDG)~\cite{PDGP}.  

On the other hand, Cosyn {\it et al.} calculate the nuclear 
transparency as a ratio of the eight fold differential cross section for pion 
electroproduction in a relativistic multiple-scattering Glauber approximation 
(RMSGA) integrated over the kinematic range of the experiment to that in a 
relativistic plane wave impulse approximation (RPWIA). In the RPWIA all 
particles are taken to be relativistic plane waves, while in the RMSGA, the 
wave function of the spectator nucleon and the outgoing pion is taken to be a 
convolution of a relativistic plane wave and a Glauber-type eikonal phase 
operator that parametrizes the effects of FSI. The parametrization chosen by 
these authors reflect the diffractive nature of nucleon-nucleon ($N' N$) and 
pion-nucleon ($\pi N$) collisions at intermediate energies. The parameters 
 $\sigma^{tot}_{iN}$ (total cross section, with $i$ as the outgoing pion or 
nucleon), $\beta_{iN}$ (slope parameter), and $\epsilon_{iN}$ (ratio of real to
 imaginary part of the scattering amplitude) were obtained by fits to the 
$N' N \rightarrow N' N$ databases from the PDG~\cite{PDGP}, for $i=N'$. 
For $i = \pi$, the parameters were obtained from fits to PDG~\cite{PDGP} 
databases, SAID~\cite{said} and Ref.~\cite{slope}. For outgoing nucleons with 
kinetic energy lower than $\sim$ 300~MeV, the Glauber formalism is no longer 
applicable and the FSI were parametrized in a relativistic optical model 
eikonal approximation (ROMEA)~\cite{romea}, with the global (S-V) optical model 
parametrization of Cooper {\it et al.}~\cite{cooper}. CT was incorporated by 
replacing the total cross section parameter$\sigma^{tot}_{iN}$ with an 
effective one based on the quantum diffusion model~\cite{Farrar}, this mirrors 
the effective interaction parameter of Larson {\it et al.}, described 
earlier. The parameters used for $l_c$ were exactly the same for both 
set of authors. Cosyn {\it et al.} also include the effects of 
short range correlations (SRC) in their calculations. The Glauber phase factor 
described above is corrected for SRC by replacing the single nucleon density 
typically used in Glauber-type calculations with an effective density. The 
effective density modifies the single nucleon density with a Jastrow 
correlation function and normalization functions which ensure the integral of 
the effective density is equal to the total number of nucleons.           
\begin{figure}
\begin{center}
\includegraphics[width=90mm,height=85mm]{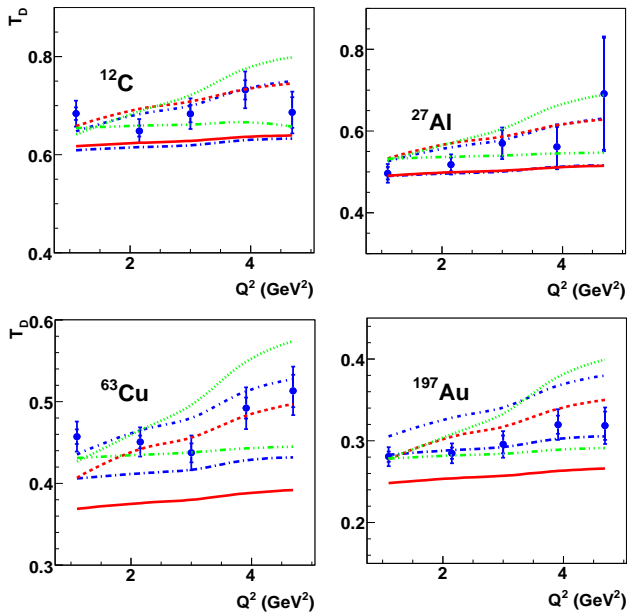}
\caption
{ \it (color online) The redefined Nuclear transparency $T_D$ versus $Q^2$ for $^{12}$C, $^{27}$Al, $^{63}$Cu, and $^{197}$Au. Curves represent calculations identical to those in Fig.~\ref{fig:nt}. Model uncertainties are identical to Fig.~\ref{fig:nt} are hence are not repeated here.}  
\label{fig:ntld2}
\end{center}
\end{figure}

We have also compared our results to the recent calculations of Kaskulov {\it et al.}, who calculate nuclear transparency as the ratio of their model 
differential cross section calculated in the lab frame, with and without FSI, 
where both types of model cross sections are integrated over the kinematic 
range of the experiment. 
Their model is built around a microscopic 
description~\cite{Mosel} of the elementary process $p(e,e'\pi^+)n$ which is 
divided into a soft hadronic and a
hard partonic or deep inelastic scattering production. For the reaction 
on nuclei, the elementary interaction is kept the same, and standard effects 
such as Fermi motion, Pauli blocking and nuclear shadowing are accounted for. 
Finally all produced pre-hadrons and hadrons are propagated through the 
nuclear medium according to the Boltzmann-Uehling-Uhlenbeck (BUU) transport 
equation. The DIS contribution to the cross section is determined by the Lund fragmentation 
model~\cite{lund} and the time development of the interactions of the 
pre-hadron is determined 
by the quantum diffusion model~\cite{Farrar}. The production time and the 
formation time are from a Monte Carlo calculation based on the Lund 
fragmentation model~\cite{lund} described in Ref.~\cite{galmc}. Only the DIS 
part of the cross section is effected by this 
pre-hadronic interaction and thus in this model only the DIS events are 
responsible for the CT effect.
            
Our results are in good agreement with the CT calculations of
Larson {\it et al.}, while the calculations of both Cosyn {\it et al.}
and Kaskulov {\it et al.} seem to somewhat overestimate the data. However,
whereas one can argue about details of the calculation, it is more important
to note that the trend of all calculations including CT in
Figs.~\ref{fig:nt} and ~\ref{fig:ntld2} are consistent with
the trend of the nuclear transparency data versus $Q^2$.

The underlying cause for the rise in nuclear transparency is different for
the different model calculations, however. Whereas the dominance of the longitudinal-photon production mechanism in exclusive (e,e$^\prime \pi^+$) reaction is thought to proceed only at asymptotic $Q^2$, and one would thus anticipate to see the CT effects enter this longitudinal channel, the Kaskulov {\it et al.} calculations find the CT effects coming from the transverse-photon production mechanism. Thus, we separately show in Fig.~\ref{fig:nt}, the nuclear transparency results for our low
$\epsilon$ data (but note that the calculations shown in Fig.~\ref{fig:ntld2}
are for the high $\epsilon$ kinematics only). Clearly, within the experimental uncertainties, we do not see any obvious difference between the nuclear 
transparency results measured at high and at low $\epsilon$ values. Given the uncertainties of the present experiment, we can not distinguish between the suggested mechanisms.

\subsection{Further studies of the CT mechanism}

Figs.~\ref{fig:nt} and ~\ref{fig:ntld2} show
 a rise of nuclear
transparency with outgoing pion momentum, or alternatively $Q^2$, which
deviates from the traditional nuclear physics expectation.
However, as can be readily witnessed from Table~\ref{table:kine},
there exists a strong correlation between the outgoing
pion momentum, $P_{\pi}$, and the magnitude of the virtual-pion
(three-)momentum, $k_{\pi}$. This poses a potential pitfall in that
the observed CT-like behavior could be an artifact due to increased values
of $k_{\pi}$ (and thus increased probability for reaction mechanisms
beyond the quasi-free picture), rather than a dependence on $P_{\pi}$. 
To investigate this further, we performed measurements at different
values of $k_{\pi}$ (and thus different $P_{\pi}$), but at identical
$Q^{2}$ (2.15 GeV$^2$).

\begin{figure}
\begin{center}
\includegraphics[width=80mm,height=75mm]{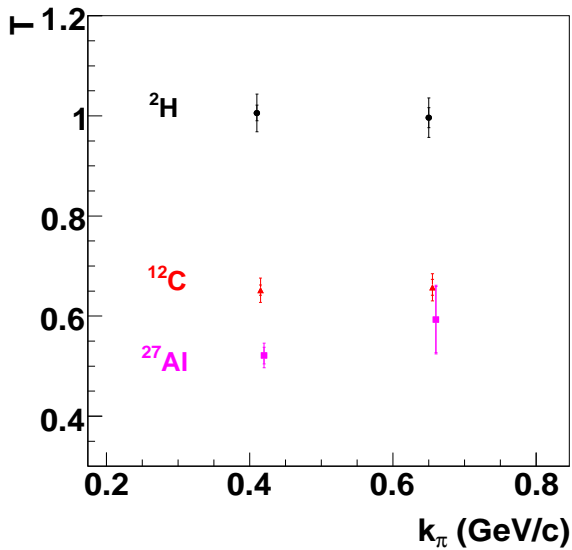}
\caption
{\it (color online) Nuclear transparency, T vs $k_{\pi}$ for $^2$H, $^{12}$C and
  $^{27}$Al. 
}  
\label{fig:ntkpi}
\end{center}
\end{figure}

The results are shown in Fig.~\ref{fig:ntkpi}, and indicate that
the nuclear transparency does not show any obvious dependence on $k_{\pi}$.
This result rules out the possibility that nuclear transparency {\sl only}
depends on $k_{\pi}$ (but does not yet rule out the possibility that
nuclear transparency depends on $k_{\pi}$ as well as other variables).

The values of $Q^2$ and $P_{\pi}$ for this exclusive (e,e$^\prime \pi^+$)
experiment are also strongly correlated. Since $Q^2$ is in general terms
related to the size of the PLC, and $P_{\pi}$ to the formation length of
the PLC, the rise in nuclear transparency results can also be a mixed
PLC size and formation-length effect. To further study which of these
two effects dominates, we formed nuclear transparency ratios of the
heavy target nuclei ($^{27}$Al, $^{63}$Cu, and $^{197}$Au) with respect
to $^{12}$C termed as $T_{C}$. This ratios should be less sensitive to formation length effects. Here we use the nuclear size as a yardstick to gauge against formation length effects.

\begin{figure}
\begin{center}
\includegraphics[width=90mm,height=85mm]{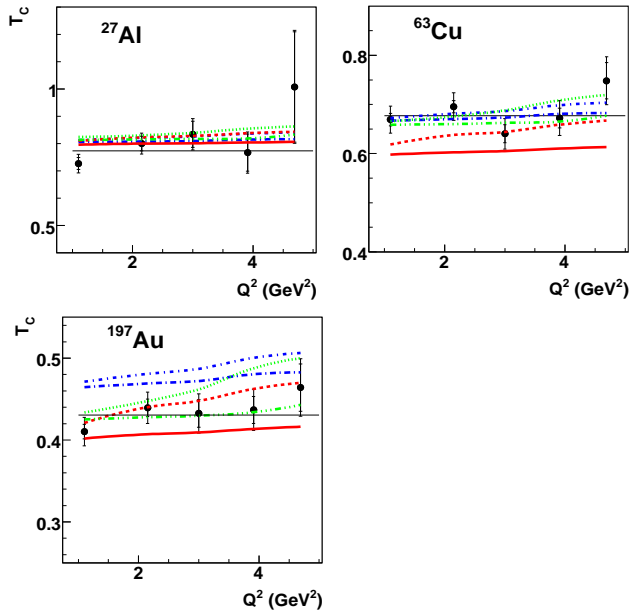}
\caption
{\it 
 (color online) The $T_{C}$ vs $Q^2$, where $T_{C}$ is defined as the super ratio of the heavy nuclear targets with respect to $^{12}$C ($T (A > 12)$ to $T (A=12)$). The inner error bars are the
 statistical uncertainties, the outer error bars are the statistical
 and point-to-point systematic uncertainties added in quadrature. The
 probabilities of the fitting to a straight line are 32\%, 40\% and
 64\% for carbon, copper and gold targets, respectively. The calculations are 
same as in Fig.~\ref{fig:nt}, but divided by the results for $^{12}$C. 
 }  
\label{fig:ntnt}
\end{center}
\end{figure}

The ``super ratios'' $T_{C}$ for $^{27}$Al (top right panel),
$^{63}$Cu (bottom left panel), and $^{197}$Au (bottom right panel)
are shown in Fig.~\ref{fig:ntnt}. The results are consistent with
a flat line, within the (large) experimental uncertainties,
with probabilities of 0.32, 0.40, and 0.64, respectively. A plausible
explanation is that the pion formation length in our kinematics is already
much larger than the nucleus radius (a simple estimate would give a formation
length of the order of 10 fm). We also note that a reasonable approximation for
the coherence length of the virtual photon ($l_c = 1/(2Mx)$) renders
a value of about 0.2-0.5 fm, which is much smaller than the size of the nucleus.
Thus, we should not be sensitive to any coherence length effects.
We conclude that these results seem to favor the explanation that the rise
in nuclear transparency measured is due to a small PLC size, rather than
a PLC formation length effect.

\subsection{The $A$ dependence}

The dependence of the nuclear transparency data on the atomic number $A$
gives further insight in the proper interpretation of the data in terms of
an onset of CT. This goes beyond the $Q^2$ or $P_{\pi}$ dependence of nuclear
transparencies for one single nuclear target described above.
Here, the entire nuclear transparency data set was examined using
a simple description with an effective parameter $\alpha$ (for each value
of $Q^2$). Using only one parameter obviously neglects specific surface
effects of the various nuclei, but has proven to be an effective way to
describe bulk properties of the nuclear medium. For example, pion-nucleus
scattering total cross section data are well described using
such a single parameter, as $\sigma^A = A^\alpha \sigma^N$, with $\sigma^A$
the nuclear cross section, $\sigma^N$ the nucleon cross section. and
$\alpha$ = 0.76~\cite{carrol1}. No noticeable dependence on the incident
pion energy was measured.

\begin{figure}
\begin{center}
\includegraphics[width=80mm,height=60mm]{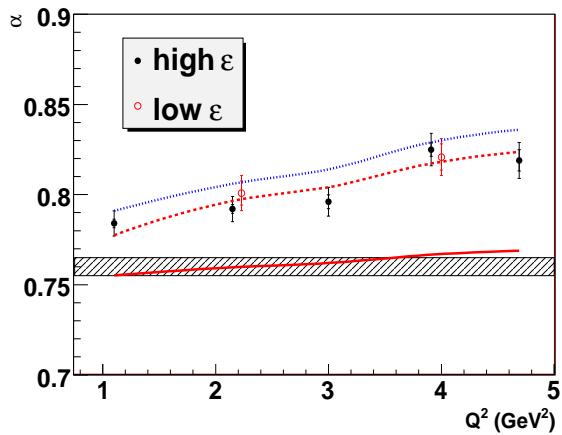} 
\caption
    {\it (color online) The parameter $\alpha$, as extracted from the global nuclear data set
    of this experiment (from $T$ = $A^{\alpha - 1)}$) versus $Q^2$
    (solid black circles). The inner error bars
    indicate the statistical uncertainty, and the outer error bars are the
    quadrature sum of statistical, systematic and modeling
    uncertainties. The hatched band is the value of $\alpha$ extracted
    from pion-nucleus scattering data~\cite{carrol1}. The solid,
    dashed and dotted lines are $\alpha$ obtained from fitting the
    $A$-dependence of the theoretical calculations: the Glauber and
    Glauber+CT calculations of Ref.~\cite{Larson06ge}, and
    the Glauber + CT (including short-range correlation effects)
    calculations of Ref.~\cite{Cosyn}, respectively.
    The red solid squares in addition show the $\alpha$
    value extracted at low $\epsilon$ value at two values of $Q^2$.
    }  
    \label{fig:alpha}
\end{center}
\end{figure}

The full nuclear transparency data set from pionCT was fitted as function
of $Q^2$ (with $T =  A^{\alpha-1}$ which equates to the same fit form as
above) are shown in Fig.~\ref{fig:alpha}. The uncertainties are
dominated by systematics, and include contributions from both fitting and
model uncertainties. The results indicate a value for the parameter
$\alpha$ clearly deviating from the total pion-nucleus scattering cross section
results, with values $\alpha >$ 0.76. Furthermore, a noticeable dependence
of $\alpha$ on $Q^2$, or equivalently the pion kinetic energy, was measured.

We compare the extracted values of $\alpha$, as function of $Q^2$, with
the calculations including CT effects of Larson {\it et al.}~\cite{Larson06ge}
and Cosyn {\it et al.}~\cite{Cosyn}. The agreement with the calculations from
Ref.~\cite{Larson06ge} is excellent, but the data are systematically below
the calculations (including both CT and short-range correlation effects)
of Ref.~\cite{Cosyn}. As mentioned above, Kaskulov {\it et
  al.}~\cite{Mosel} recently suggested that the CT effect should only
exist in the transverse 
cross section. For this reason, we have again separately indicated the $\alpha$
values for the low and high $\epsilon$ values of this experiment, where
applicable. Within our uncertainties, we see no indication of this prediction.
However, we have to warn that the difference between the low- to high-$\epsilon$
kinematics in terms of contributions from the longitudinal cross sections to
the total measured cross sections changes by less than 30\% in our kinematics
(see Fig.~\ref{fig:qs} and Table~\ref{table:kine}). It is thus very well
possible that the measured effect is solely due to the transverse
contributions.

\section{Conclusions}
\label{sec:concl}

The $A(e,e^\prime \pi^+)$ reaction was studied on a variety of nuclear
targets: $^1$H, $^2$H, $^{12}$C, $^{27}$Al, Cu, and Au. Data were
taken up to a four-momentum transfer squared of $Q^2$ = 4.8 GeV$^2$,
and analyzed in terms of nuclear transparencies, the escape probability
of the positively-charged pion from the nuclear medium. A rise of the
nuclear transparency with $Q^2$ or pion momentum could signal an
onset of Color Transparency, expected to occur at large
values of $Q^2$ from both perturbative and non-perturbative QCD.

The existence of Color Transparency is well-known. The most convincing evidence
is the analysis of Fermilab E791 data on the diffractive dissociation in two
jets. However, conclusive proof of the {\sl onset} of CT
is more elusive, although hints of it were recently seen in the analysis of
$\rho$ vector meson data at fixed coherence length~\cite{Airape} and
$\pi^-$ photo-production~\cite{Dutta}. Our results show a far more
conclusive onset of CT over the relatively large range in pion momentum 
between 2.5 and 4.5
GeV/c, and the $Q^2$ range between 1.1 and 4.8 GeV$^2$, and are in
good agreement with model calculations including the CT effect of
Larson {\it et al.}, Cosyn {\it et al.}, and Kaskulov {\it et al.}
~\cite{Larson06ge,Cosyn,mosel2}. The latter calculation also agrees
excellently with the measured $-t$ dependence of the differential cross 
sections, giving further credence to the onset of CT.

Specialized data sets were added to ensure that the noted rise of nuclear
transparencies is indeed due to CT. In particular, the cross checks performed
are:
\begin{enumerate}
\item {L/T character of the cross section\\
Within uncertainties, we find the longitudinal-transverse character of
the pion electroproduction cross sections to be similar off protons and
heavier nuclei. This supports a quasifree reaction mechanism.}
\item {Virtual-pion momentum $k_\pi$ \\
The measured nuclear transparencies do not appear to depend on the
virtual-pion (three-)momentum, which could be a signal for increased
reaction mechanism effects beyond the quasifree picture. This suggests
that reaction mechanism effects beyond the quasifree are suppressed.}
\item {Cross section ratios of medium- to heavy-nuclei\\
``Super ratios'' of the heavier target nuclei ($^{27}$Al, Cu, Au) with respect
to $^{12}$C show a similar rise in nuclear transparencies, indicating that
the pion formation length in our kinematics is already far longer than the
nuclear radius.}
\item {$A$ dependence \\
The $A$ dependence of the data can be described by a single parameter, 
$\alpha$. It was found to be consistently larger than the 0.76 found from 
pion-nucleus total cross section data, {\sl and} rising with $Q^2$, consistent 
with a CT Ansatz.}
\end{enumerate}
Furthermore, our results are at nearly-constant and small coherence length
($\approx$ 0.4 fm), such that possible complications due to $\pi\rho$
exchange terms are minimized.

Our results are consistent with the predicted early onset of CT in
mesons compared to baryons, and together with previous meson transparency
measurements~\cite{Airape,Dutta} suggest a gradual transition to meson
production with small inter-quark separation. These results put severe
constraints on early models of CT which predict a dramatic transition with
a threshold-like behavior. The unambiguous observation of the onset of
CT uniquely points to the role of color in exclusive
high-$Q^2$ processes. Furthermore, it is an effective signature of the 
approach to
the factorization regime in meson electroproduction experiments, necessary
for the access to Generalized Parton Distributions. These results will be
further extended by data to be taken after the Jefferson Lab upgrade to 
12 GeV, with planned exclusive $A(e,e^\prime \pi^+)$ measurements up to
$Q^2 \approx$ 10 GeV$^2$~\cite{E1206107}.

\section{Acknowledgments}
The authors would like to acknowledge the outstanding support of Jefferson Lab Hall C 
technical staffs and Accelerator Division in accomplishing this experiment. 
This work was supported 
in part by the U.S. Department of Energy. The Southeastern Universities 
Research Association (SURA) operates the Thomas Jefferson National Accelerator 
Facility for the United States Department of Energy under contract 
DE-AC05-84150. We acknowledge additional research grants from the U.S. 
National Science Foundation, the Natural Sciences and Engineering Research 
Council of Canada (NSERC), NATO, FOM (Netherlands), and KOSEF (South Korea).

\hyphenation{Post-Script Sprin-ger}

\begin{table*}
{\bf{\large Appendix A}}
  \begin{small}
  \begin{center}
    \begin{tabular}{|c|c|c|c|c|c|c|c|c|c|}
      \hline
      Target & $Q^2$ & W & -t & $P_{\pi}$ & $\epsilon$ & $P^{CM}_{\pi}$  &
      $\frac{d\sigma}{dtdP^{CM}_{\pi}}$ & stat. err. & sys. err.\\\hline
        & GeV$^2$ & GeV &  GeV$^2$ & GeV/c &  & GeV/c  & $\mu b/ GeV^3$& 
      $\mu b/ GeV^3$ &
      $\mu b/ GeV^3$ \\\hline
$LH_2$ & 1.71 & 2.36 & 0.100 & 3.35 & 0.532 & 0.99 & 3.72 & 0.11 & 0.24 \\ \hline
$LH_2$ & 1.80 & 2.33 & 0.117 & 3.31 & 0.538 & 0.97 & 3.46 & 0.11 & 0.23 \\ \hline
$LH_2$ & 1.84 & 2.32 & 0.128 & 3.30 & 0.539 & 0.96 & 3.47 & 0.11 & 0.20 \\ \hline
$LH_2$ & 1.87 & 2.31 & 0.136 & 3.30 & 0.538 & 0.96 & 3.04 & 0.10 & 0.19 \\ \hline
$LH_2$ & 1.90 & 2.30 & 0.144 & 3.26 & 0.545 & 0.95 & 3.07 & 0.10 & 0.19 \\ \hline
$LH_2$ & 1.93 & 2.28 & 0.152 & 3.24 & 0.548 & 0.94 & 2.92 & 0.09 & 0.17 \\ \hline
$LH_2$ & 1.94 & 2.27 & 0.160 & 3.21 & 0.554 & 0.93 & 2.96 & 0.09 & 0.18 \\ \hline
$LH_2$ & 1.95 & 2.27 & 0.167 & 3.21 & 0.555 & 0.93 & 2.78 & 0.09 & 0.17 \\ \hline
$LH_2$ & 2.02 & 2.25 & 0.174 & 3.20 & 0.554 & 0.92 & 2.75 & 0.08 & 0.16 \\ \hline
$LH_2$ & 2.06 & 2.24 & 0.181 & 3.19 & 0.553 & 0.91 & 2.54 & 0.08 & 0.15 \\ \hline
$LH_2$ & 2.05 & 2.24 & 0.187 & 3.19 & 0.553 & 0.91 & 2.71 & 0.09 & 0.15 \\ \hline
$LH_2$ & 2.09 & 2.23 & 0.193 & 3.17 & 0.555 & 0.90 & 2.61 & 0.09 & 0.14 \\ \hline
$LH_2$ & 2.08 & 2.24 & 0.200 & 3.17 & 0.555 & 0.91 & 2.48 & 0.07 & 0.13 \\ \hline
$LH_2$ & 2.09 & 2.23 & 0.208 & 3.17 & 0.554 & 0.90 & 2.38 & 0.07 & 0.12 \\ \hline
$LH_2$ & 2.13 & 2.23 & 0.216 & 3.16 & 0.552 & 0.90 & 2.55 & 0.07 & 0.13 \\ \hline
$LH_2$ & 2.13 & 2.23 & 0.224 & 3.16 & 0.552 & 0.90 & 2.27 & 0.07 & 0.11 \\ \hline
$LH_2$ & 2.13 & 2.23 & 0.232 & 3.16 & 0.552 & 0.90 & 2.29 & 0.07 & 0.11 \\ \hline
$LH_2$ & 2.20 & 2.19 & 0.240 & 3.10 & 0.563 & 0.88 & 2.21 & 0.07 & 0.11 \\ \hline
$LH_2$ & 2.22 & 2.19 & 0.249 & 3.10 & 0.560 & 0.87 & 2.18 & 0.06 & 0.11 \\ \hline
$LH_2$ & 2.19 & 2.19 & 0.259 & 3.10 & 0.563 & 0.88 & 2.15 & 0.07 & 0.11 \\ \hline
$LH_2$ & 2.26 & 2.18 & 0.270 & 3.08 & 0.558 & 0.86 & 2.11 & 0.06 & 0.11 \\ \hline
$LH_2$ & 2.29 & 2.17 & 0.283 & 3.08 & 0.557 & 0.86 & 2.05 & 0.06 & 0.12 \\ \hline
$LH_2$ & 2.34 & 2.16 & 0.299 & 3.07 & 0.557 & 0.86 & 1.95 & 0.06 & 0.12 \\ \hline
$LH_2$ & 2.44 & 2.13 & 0.317 & 3.06 & 0.556 & 0.84 & 2.11 & 0.07 & 0.15 \\ \hline
$LH_2$ & 2.53 & 2.12 & 0.341 & 3.04 & 0.547 & 0.82 & 1.78 & 0.06 & 0.04 \\ \hline
$LH_2$ & 2.63 & 2.08 & 0.385 & 3.03 & 0.561 & 0.82 & 1.83 & 0.07 & 0.19 \\ \hline
$LH_2$ & 3.75 & 2.21 & 0.425 & 3.89 & 0.251 & 0.89 & 0.64 & 0.02 & 0.06 \\ \hline
$LH_2$ & 4.00 & 2.14 & 0.548 & 3.80 & 0.255 & 0.85 & 0.58 & 0.02 & 0.06 \\ \hline
$LH_2$ & 4.00 & 2.14 & 0.646 & 3.73 & 0.255 & 0.84 & 0.64 & 0.02 & 0.07 \\ \hline
$LH_2$ & 4.21 & 2.07 & 0.758 & 3.63 & 0.261 & 0.80 & 0.70 & 0.02 & 0.07 \\ \hline
$LH_2$ & 2.55 & 2.28 & 0.199 & 3.55 & 0.435 & 0.94 & 1.82 & 0.06 & 0.09 \\ \hline
$LH_2$ & 2.64 & 2.26 & 0.236 & 3.52 & 0.433 & 0.93 & 1.70 & 0.05 & 0.09 \\ \hline
$LH_2$ & 2.68 & 2.24 & 0.260 & 3.49 & 0.438 & 0.91 & 1.50 & 0.05 & 0.08 \\ \hline
$LH_2$ & 2.72 & 2.23 & 0.280 & 3.46 & 0.441 & 0.91 & 1.41 & 0.05 & 0.07 \\ \hline
$LH_2$ & 2.78 & 2.21 & 0.300 & 3.45 & 0.440 & 0.90 & 1.43 & 0.05 & 0.08 \\ \hline
$LH_2$ & 2.85 & 2.19 & 0.320 & 3.42 & 0.443 & 0.88 & 1.42 & 0.04 & 0.08 \\ \hline
$LH_2$ & 2.90 & 2.18 & 0.340 & 3.40 & 0.445 & 0.87 & 1.26 & 0.04 & 0.07 \\ \hline
$LH_2$ & 2.94 & 2.16 & 0.360 & 3.38 & 0.445 & 0.86 & 1.36 & 0.04 & 0.07 \\ \hline
$LH_2$ & 2.96 & 2.16 & 0.380 & 3.37 & 0.446 & 0.86 & 1.28 & 0.04 & 0.07 \\ \hline
$LH_2$ & 3.06 & 2.13 & 0.400 & 3.33 & 0.450 & 0.84 & 1.26 & 0.04 & 0.07 \\ \hline
$LH_2$ & 3.08 & 2.12 & 0.420 & 3.30 & 0.454 & 0.83 & 1.22 & 0.04 & 0.06 \\ \hline
$LH_2$ & 3.11 & 2.11 & 0.445 & 3.29 & 0.454 & 0.82 & 1.16 & 0.03 & 0.06 \\ \hline
$LH_2$ & 3.27 & 2.07 & 0.474 & 3.28 & 0.450 & 0.80 & 1.08 & 0.04 & 0.06 \\ \hline
$LH_2$ & 3.25 & 2.08 & 0.508 & 3.25 & 0.447 & 0.80 & 1.09 & 0.04 & 0.06 \\ \hline
$LH_2$ & 3.50 & 1.99 & 0.565 & 3.21 & 0.455 & 0.77 & 1.22 & 0.04 & 0.07 \\ \hline
$LH_2$ & 0.92 & 2.31 & 0.038 & 2.84 & 0.490 & 0.96 & 10.74 & 0.32 & 0.53 \\ \hline
$LH_2$ & 0.98 & 2.29 & 0.047 & 2.81 & 0.497 & 0.94 & 9.56 & 0.26 & 0.49 \\ \hline
$LH_2$ & 1.00 & 2.28 & 0.053 & 2.80 & 0.499 & 0.94 & 8.99 & 0.26 & 0.46 \\ \hline
\end{tabular}
  \end{center}
\end{small}
\caption{Extracted cross sections and their uncertainties for hydrogen
  target at eight kinematics 
  settings. The $Q^2$, W, $-t$, $P_{\pi}$ and $\epsilon$ values for
  each kinematics are shown at the same time. } 
\label{table:h1dxs}
\end{table*}

\begin{table*}
  \begin{small}
  \begin{center}
    \begin{tabular}{|c|c|c|c|c|c|c|c|c|c|}
      \hline
      Target & $Q^2$ & W & -t & $P_{\pi}$  & $\epsilon$ & $P^{CM}_{\pi}$  &
      $\frac{d\sigma}{dtdP^{CM}_{\pi}}$ & stat. err. & sys. err.\\\hline
        & GeV$^2$ & GeV &  GeV$^2$ & GeV/c & & GeV/c  & $\mu b/ GeV^3$& 
      $\mu b/ GeV^3$ &
      $\mu b/ GeV^3$ \\\hline
$LH_2$ & 1.04 & 2.27 & 0.058 & 2.79 & 0.501 & 0.93 & 8.15 & 0.22 & 0.42 \\ \hline
$LH_2$ & 1.03 & 2.27 & 0.062 & 2.78 & 0.501 & 0.93 & 7.87 & 0.23 & 0.42 \\ \hline
$LH_2$ & 1.10 & 2.26 & 0.066 & 2.79 & 0.497 & 0.92 & 7.90 & 0.22 & 0.42 \\ \hline
$LH_2$ & 1.07 & 2.26 & 0.070 & 2.76 & 0.504 & 0.92 & 7.35 & 0.21 & 0.39 \\ \hline
$LH_2$ & 1.10 & 2.25 & 0.074 & 2.77 & 0.502 & 0.92 & 6.73 & 0.19 & 0.35 \\ \hline
$LH_2$ & 1.10 & 2.25 & 0.078 & 2.76 & 0.501 & 0.92 & 7.06 & 0.19 & 0.37 \\ \hline
$LH_2$ & 1.12 & 2.24 & 0.082 & 2.75 & 0.503 & 0.91 & 6.71 & 0.19 & 0.34 \\ \hline
$LH_2$ & 1.13 & 2.24 & 0.086 & 2.74 & 0.506 & 0.91 & 6.52 & 0.19 & 0.33 \\ \hline
$LH_2$ & 1.16 & 2.23 & 0.090 & 2.74 & 0.504 & 0.91 & 6.34 & 0.17 & 0.32 \\ \hline
$LH_2$ & 1.16 & 2.24 & 0.095 & 2.73 & 0.503 & 0.91 & 6.00 & 0.17 & 0.30 \\ \hline
$LH_2$ & 1.19 & 2.22 & 0.100 & 2.71 & 0.510 & 0.89 & 6.58 & 0.19 & 0.33 \\ \hline
$LH_2$ & 1.20 & 2.22 & 0.105 & 2.72 & 0.505 & 0.90 & 6.01 & 0.19 & 0.31 \\ \hline
$LH_2$ & 1.21 & 2.22 & 0.110 & 2.71 & 0.505 & 0.89 & 6.50 & 0.20 & 0.33 \\ \hline
$LH_2$ & 1.18 & 2.23 & 0.116 & 2.70 & 0.506 & 0.90 & 6.18 & 0.19 & 0.32 \\ \hline
$LH_2$ & 1.16 & 2.24 & 0.123 & 2.71 & 0.501 & 0.90 & 5.51 & 0.18 & 0.30 \\ \hline
$LH_2$ & 1.17 & 2.23 & 0.133 & 2.69 & 0.505 & 0.89 & 6.15 & 0.18 & 0.37 \\ \hline
$LH_2$ & 1.21 & 2.22 & 0.154 & 2.67 & 0.502 & 0.88 & 7.21 & 0.20 & 0.59 \\ \hline
$LH_2$ & 2.06 & 2.24 & 0.165 & 3.21 & 0.268 & 0.91 & 2.52 & 0.08 & 0.12 \\ \hline
$LH_2$ & 2.16 & 2.21 & 0.209 & 3.17 & 0.269 & 0.90 & 2.01 & 0.07 & 0.10 \\ \hline
$LH_2$ & 2.26 & 2.18 & 0.248 & 3.12 & 0.274 & 0.88 & 1.95 & 0.06 & 0.10 \\ \hline
$LH_2$ & 2.27 & 2.18 & 0.290 & 3.07 & 0.276 & 0.86 & 1.97 & 0.06 & 0.10 \\ \hline
$LH_2$ & 2.28 & 2.18 & 0.326 & 3.02 & 0.274 & 0.85 & 2.04 & 0.07 & 0.13 \\ \hline
$LH_2$ & 1.81 & 1.86 & 0.241 & 2.21 & 0.634 & 0.68 & 5.74 & 0.19 & 0.43 \\ \hline
$LH_2$ & 1.86 & 1.84 & 0.276 & 2.17 & 0.634 & 0.67 & 5.01 & 0.16 & 0.33 \\ \hline
$LH_2$ & 1.93 & 1.82 & 0.305 & 2.15 & 0.633 & 0.65 & 5.10 & 0.15 & 0.31 \\ \hline
$LH_2$ & 2.00 & 1.79 & 0.330 & 2.12 & 0.636 & 0.63 & 5.20 & 0.18 & 0.31 \\ \hline
$LH_2$ & 2.08 & 1.76 & 0.355 & 2.09 & 0.638 & 0.62 & 5.59 & 0.15 & 0.32 \\ \hline
$LH_2$ & 2.14 & 1.74 & 0.385 & 2.07 & 0.637 & 0.60 & 5.61 & 0.15 & 0.32 \\ \hline
$LH_2$ & 2.19 & 1.74 & 0.414 & 2.07 & 0.629 & 0.60 & 5.73 & 0.16 & 0.33 \\ \hline
$LH_2$ & 2.25 & 1.71 & 0.444 & 2.04 & 0.630 & 0.58 & 5.58 & 0.18 & 0.30 \\ \hline
$LH_2$ & 2.30 & 1.70 & 0.479 & 2.02 & 0.628 & 0.57 & 5.61 & 0.19 & 0.30 \\ \hline
$LH_2$ & 2.40 & 1.67 & 0.548 & 1.98 & 0.622 & 0.55 & 5.58 & 0.17 & 0.29 \\ \hline
$LH_2$ & 4.39 & 2.32 & 0.469 & 4.49 & 0.259 & 0.96 & 0.45 & 0.02 & 0.02 \\ \hline
$LH_2$ & 4.71 & 2.24 & 0.621 & 4.37 & 0.265 & 0.91 & 0.41 & 0.01 & 0.02 \\ \hline
$LH_2$ & 4.82 & 2.21 & 0.765 & 4.28 & 0.269 & 0.90 & 0.41 & 0.01 & 0.02 \\ \hline
$LH_2$ & 4.98 & 2.17 & 0.914 & 4.17 & 0.267 & 0.86 & 0.43 & 0.01 & 0.02 \\ \hline
$LH_2$ & 3.45 & 2.38 & 0.280 & 4.24 & 0.381 & 1.00 & 0.75 & 0.02 & 0.04 \\ \hline
$LH_2$ & 3.55 & 2.36 & 0.340 & 4.20 & 0.381 & 0.99 & 0.71 & 0.02 & 0.04 \\ \hline
$LH_2$ & 3.69 & 2.32 & 0.380 & 4.17 & 0.383 & 0.97 & 0.67 & 0.02 & 0.04 \\ \hline
$LH_2$ & 3.78 & 2.30 & 0.420 & 4.12 & 0.386 & 0.95 & 0.68 & 0.02 & 0.04 \\ \hline
$LH_2$ & 3.84 & 2.28 & 0.460 & 4.09 & 0.386 & 0.93 & 0.62 & 0.02 & 0.03 \\ \hline
$LH_2$ & 3.91 & 2.26 & 0.499 & 4.05 & 0.388 & 0.92 & 0.59 & 0.02 & 0.03 \\ \hline
$LH_2$ & 3.92 & 2.25 & 0.540 & 4.00 & 0.393 & 0.91 & 0.62 & 0.02 & 0.03 \\ \hline
$LH_2$ & 4.00 & 2.23 & 0.579 & 3.97 & 0.395 & 0.90 & 0.63 & 0.02 & 0.03 \\ \hline
$LH_2$ & 4.10 & 2.21 & 0.619 & 3.93 & 0.395 & 0.88 & 0.61 & 0.02 & 0.03 \\ \hline
$LH_2$ & 4.30 & 2.15 & 0.663 & 3.91 & 0.394 & 0.85 & 0.66 & 0.02 & 0.03 \\ \hline
$LH_2$ & 4.40 & 2.12 & 0.741 & 3.87 & 0.398 & 0.84 & 0.63 & 0.02 & 0.03 \\ \hline
    \end{tabular}
  \end{center}
\end{small}
\caption{Continue of Table.~\ref{table:h1dxs}} 
\label{table:h2dxs}
\end{table*}

\begin{table*}
  {\bf{\large Appendix B}}
  \begin{small}
  \begin{center}
    \begin{tabular}{|c|c|c|c|c|c|c|c|c|c|}
      \hline
      Target & $Q^2$ & W & -t & $P_{\pi}$  & $P^{CM}_{\pi}$ & $\epsilon$ &
      $\frac{d\sigma}{dtdP^{CM}_{\pi}}$ & stat. err. & sys. err.\\\hline
        & GeV$^2$ & GeV &  GeV$^2$ & GeV/c & GeV/c &  & $\mu b/ GeV^3$& 
      $\mu b/ GeV^3$ &
      $\mu b/ GeV^3$ \\\hline
$LD_2$ & 2.24 & 3.68 & 0.177 & 3.20 & 1.35 & 0.559 & 17.36 & 0.63 & 1.30 \\ \hline
$LD_2$ & 2.24 & 3.68 & 0.188 & 3.19 & 1.34 & 0.559 & 31.72 & 1.16 & 1.82 \\ \hline
$LD_2$ & 2.24 & 3.68 & 0.195 & 3.18 & 1.34 & 0.559 & 45.58 & 1.73 & 2.60 \\ \hline
$LD_2$ & 2.24 & 3.68 & 0.203 & 3.17 & 1.34 & 0.559 & 55.31 & 2.03 & 3.10 \\ \hline
$LD_2$ & 2.24 & 3.68 & 0.211 & 3.16 & 1.33 & 0.559 & 64.42 & 2.48 & 3.72 \\ \hline
$LD_2$ & 2.24 & 3.68 & 0.220 & 3.15 & 1.33 & 0.559 & 66.01 & 2.42 & 3.98 \\ \hline
$LD_2$ & 2.24 & 3.68 & 0.229 & 3.14 & 1.32 & 0.559 & 59.77 & 2.20 & 3.78 \\ \hline
$LD_2$ & 2.24 & 3.68 & 0.240 & 3.13 & 1.32 & 0.559 & 53.28 & 2.00 & 3.51 \\ \hline
$LD_2$ & 2.24 & 3.68 & 0.251 & 3.12 & 1.31 & 0.559 & 41.23 & 1.55 & 2.78 \\ \hline
$LD_2$ & 2.24 & 3.68 & 0.263 & 3.10 & 1.31 & 0.559 & 27.73 & 1.03 & 1.86 \\ \hline
$LD_2$ & 2.24 & 3.68 & 0.277 & 3.08 & 1.30 & 0.559 & 15.29 & 0.61 & 1.01 \\ \hline
$LD_2$ & 2.24 & 3.68 & 0.299 & 3.04 & 1.28 & 0.559 & 7.86 & 0.28 & 0.46 \\ \hline
$LD_2$ & 4.00 & 3.86 & 0.442 & 3.90 & 1.43 & 0.254 & 4.65 & 0.21 & 0.58 \\ \hline
$LD_2$ & 4.00 & 3.86 & 0.512 & 3.84 & 1.41 & 0.254 & 10.20 & 0.43 & 1.25 \\ \hline
$LD_2$ & 3.02 & 3.73 & 0.290 & 3.46 & 1.37 & 0.449 & 6.75 & 0.25 & 0.47 \\ \hline
$LD_2$ & 3.02 & 3.73 & 0.316 & 3.42 & 1.36 & 0.449 & 17.55 & 0.66 & 1.13 \\ \hline
$LD_2$ & 3.02 & 3.73 & 0.337 & 3.40 & 1.35 & 0.449 & 24.30 & 0.92 & 1.49 \\ \hline
$LD_2$ & 3.02 & 3.73 & 0.359 & 3.38 & 1.34 & 0.449 & 25.32 & 0.98 & 1.50 \\ \hline
$LD_2$ & 3.02 & 3.73 & 0.380 & 3.36 & 1.33 & 0.449 & 22.72 & 1.00 & 1.28 \\ \hline
$LD_2$ & 3.02 & 3.73 & 0.402 & 3.34 & 1.32 & 0.449 & 16.21 & 0.65 & 0.91 \\ \hline
$LD_2$ & 1.12 & 3.60 & 0.067 & 2.78 & 1.30 & 0.504 & 188.27 & 6.96 & 11.29 \\ \hline
$LD_2$ & 1.12 & 3.60 & 0.071 & 2.78 & 1.30 & 0.504 & 262.48 & 9.75 & 14.97 \\ \hline
$LD_2$ & 1.12 & 3.60 & 0.077 & 2.77 & 1.29 & 0.504 & 310.24 & 11.76 & 18.10 \\ \hline
$LD_2$ & 1.12 & 3.60 & 0.082 & 2.76 & 1.29 & 0.504 & 316.58 & 12.18 & 19.00 \\ \hline
$LD_2$ & 1.12 & 3.60 & 0.088 & 2.75 & 1.29 & 0.504 & 268.56 & 10.35 & 16.58 \\ \hline
$LD_2$ & 1.12 & 3.60 & 0.094 & 2.75 & 1.29 & 0.504 & 211.69 & 7.88 & 13.73 \\ \hline
$LD_2$ & 1.12 & 3.60 & 0.099 & 2.74 & 1.28 & 0.504 & 144.49 & 5.42 & 9.61 \\ \hline
$LD_2$ & 1.12 & 3.60 & 0.105 & 2.72 & 1.27 & 0.504 & 75.80 & 3.03 & 4.91 \\ \hline
$LD_2$ & 1.12 & 3.60 & 0.111 & 2.70 & 1.27 & 0.504 & 36.38 & 1.60 & 2.26 \\ \hline
$LD_2$ & 1.12 & 3.60 & 0.120 & 2.68 & 1.25 & 0.504 & 16.95 & 0.98 & 0.95 \\ \hline
$LD_2$ & 1.12 & 3.60 & 0.128 & 2.66 & 1.24 & 0.504 & 9.92 & 0.69 & 0.54 \\ \hline
$LD_2$ & 2.24 & 3.68 & 0.193 & 3.19 & 1.34 & 0.274 & 36.94 & 1.41 & 3.80 \\ \hline
$LD_2$ & 2.24 & 3.68 & 0.221 & 3.15 & 1.33 & 0.274 & 58.50 & 2.38 & 7.07 \\ \hline
$LD_2$ & 2.24 & 3.68 & 0.248 & 3.12 & 1.32 & 0.274 & 40.19 & 1.56 & 4.88 \\ \hline
$LD_2$ & 2.24 & 3.68 & 0.293 & 3.06 & 1.29 & 0.274 & 8.69 & 0.31 & 1.05 \\ \hline
$LD_2$ & 2.14 & 3.17 & 0.314 & 2.17 & 0.99 & 0.630 & 24.32 & 0.98 & 1.78 \\ \hline
$LD_2$ & 2.14 & 3.17 & 0.346 & 2.14 & 0.98 & 0.630 & 60.74 & 2.25 & 4.07 \\ \hline
$LD_2$ & 2.14 & 3.17 & 0.367 & 2.11 & 0.97 & 0.630 & 83.29 & 3.10 & 5.98 \\ \hline
$LD_2$ & 2.14 & 3.17 & 0.385 & 2.09 & 0.96 & 0.630 & 89.87 & 3.22 & 7.08 \\ \hline
$LD_2$ & 2.14 & 3.17 & 0.404 & 2.07 & 0.95 & 0.630 & 66.90 & 2.65 & 5.57 \\ \hline
$LD_2$ & 2.14 & 3.17 & 0.424 & 2.05 & 0.94 & 0.630 & 48.89 & 1.85 & 4.44 \\ \hline
$LD_2$ & 2.14 & 3.17 & 0.451 & 2.02 & 0.92 & 0.630 & 27.91 & 1.18 & 2.74 \\ \hline
$LD_2$ & 2.14 & 3.17 & 0.481 & 1.98 & 0.91 & 0.630 & 18.00 & 0.93 & 1.99 \\ \hline
$LD_2$ & 4.74 & 4.06 & 0.507 & 4.48 & 1.55 & 0.264 & 2.76 & 0.10 & 0.34 \\ \hline
$LD_2$ & 4.74 & 4.06 & 0.592 & 4.40 & 1.52 & 0.264 & 5.54 & 0.22 & 0.67 \\ \hline
$LD_2$ & 4.74 & 4.06 & 0.668 & 4.34 & 1.50 & 0.264 & 5.31 & 0.22 & 0.65 \\ \hline
$LD_2$ & 3.94 & 3.98 & 0.359 & 4.18 & 1.52 & 0.391 & 2.47 & 0.09 & 0.27 \\ \hline
$LD_2$ & 3.94 & 3.98 & 0.393 & 4.15 & 1.51 & 0.391 & 6.20 & 0.23 & 0.57 \\ \hline
$LD_2$ & 3.94 & 3.98 & 0.417 & 4.12 & 1.50 & 0.391 & 9.38 & 0.35 & 0.77 \\ \hline
$LD_2$ & 3.94 & 3.98 & 0.442 & 4.10 & 1.49 & 0.391 & 11.94 & 0.44 & 0.95 \\ \hline
$LD_2$ & 3.94 & 3.98 & 0.469 & 4.08 & 1.48 & 0.391 & 10.97 & 0.42 & 0.89 \\ \hline
$LD_2$ & 3.94 & 3.98 & 0.496 & 4.05 & 1.48 & 0.391 & 10.04 & 0.41 & 0.85 \\ \hline
$LD_2$ & 3.94 & 3.98 & 0.525 & 4.03 & 1.47 & 0.391 & 8.22 & 0.31 & 0.76 \\ \hline
\end{tabular}
  \end{center}
\end{small}
\caption{Extracted cross sections and their uncertainties for deuterium
  target at eight kinematics 
  settings. The $Q^2$, W, $-t$, $P_{\pi}$ and $\epsilon$ values for
  each kinematics are shown at the same time. } 
\label{table:ddxs}
\end{table*}

\begin{table*}
  \begin{small}
  \begin{center}
    \begin{tabular}{|c|c|c|c|c|c|c|c|c|c|}
      \hline
      Target & $Q^2$ & W & -t & $P_{\pi}$  & $P^{CM}_{\pi}$ & $\epsilon$ &
      $\frac{d\sigma}{dtdP^{CM}_{\pi}}$ & stat. err. & sys. err.\\\hline
        & GeV$^2$ & GeV &  GeV$^2$ & GeV/c & GeV/c &  & $\mu b/ GeV^3$& 
      $\mu b/ GeV^3$ &
      $\mu b/ GeV^3$ \\\hline
$C$ & 2.24 & 14.01 & 0.159 & 3.23 & 2.51 & 0.558 & 26.45 & 1.02 & 2.02 \\ \hline
$C$ & 2.24 & 14.01 & 0.177 & 3.21 & 2.49 & 0.558 & 41.07 & 1.61 & 2.94 \\ \hline
$C$ & 2.24 & 14.01 & 0.189 & 3.20 & 2.48 & 0.558 & 50.06 & 1.76 & 3.26 \\ \hline
$C$ & 2.24 & 14.01 & 0.200 & 3.18 & 2.47 & 0.558 & 54.14 & 2.09 & 3.35 \\ \hline
$C$ & 2.24 & 14.01 & 0.209 & 3.17 & 2.46 & 0.558 & 49.20 & 1.88 & 3.28 \\ \hline
$C$ & 2.24 & 14.01 & 0.218 & 3.16 & 2.45 & 0.558 & 57.30 & 2.12 & 3.51 \\ \hline
$C$ & 2.24 & 14.01 & 0.227 & 3.14 & 2.44 & 0.558 & 53.93 & 2.01 & 3.37 \\ \hline
$C$ & 2.24 & 14.01 & 0.237 & 3.13 & 2.43 & 0.558 & 56.09 & 2.08 & 3.35 \\ \hline
$C$ & 2.24 & 14.01 & 0.246 & 3.12 & 2.42 & 0.558 & 51.33 & 1.92 & 3.17 \\ \hline
$C$ & 2.24 & 14.01 & 0.257 & 3.10 & 2.41 & 0.558 & 47.62 & 1.77 & 2.95 \\ \hline
$C$ & 2.24 & 14.01 & 0.271 & 3.08 & 2.39 & 0.558 & 42.88 & 1.48 & 2.84 \\ \hline
$C$ & 4.04 & 14.59 & 0.388 & 3.96 & 2.91 & 0.257 & 4.79 & 0.32 & 0.61 \\ \hline
$C$ & 4.04 & 14.59 & 0.498 & 3.85 & 2.83 & 0.257 & 7.08 & 0.36 & 0.91 \\ \hline
$C$ & 3.02 & 14.22 & 0.261 & 3.49 & 2.65 & 0.450 & 14.12 & 0.55 & 1.77 \\ \hline
$C$ & 3.02 & 14.22 & 0.315 & 3.43 & 2.61 & 0.450 & 20.13 & 0.66 & 2.57 \\ \hline
$C$ & 3.02 & 14.22 & 0.367 & 3.37 & 2.56 & 0.450 & 22.06 & 0.69 & 2.78 \\ \hline
$C$ & 1.12 & 13.67 & 0.071 & 2.78 & 2.24 & 0.504 & 95.54 & 3.80 & 9.86 \\ \hline
$C$ & 1.12 & 13.67 & 0.079 & 2.77 & 2.23 & 0.504 & 231.96 & 8.73 & 25.10 \\ \hline
$C$ & 1.12 & 13.67 & 0.085 & 2.76 & 2.22 & 0.504 & 253.64 & 9.85 & 29.19 \\ \hline
$C$ & 1.12 & 13.67 & 0.089 & 2.75 & 2.21 & 0.504 & 264.99 & 9.82 & 31.33 \\ \hline
$C$ & 1.12 & 13.67 & 0.094 & 2.74 & 2.20 & 0.504 & 256.93 & 10.00 & 30.78 \\ \hline
$C$ & 1.12 & 13.67 & 0.099 & 2.73 & 2.19 & 0.504 & 248.65 & 9.55 & 30.26 \\ \hline
$C$ & 1.12 & 13.67 & 0.104 & 2.72 & 2.19 & 0.504 & 229.94 & 9.41 & 26.27 \\ \hline
$C$ & 1.12 & 13.67 & 0.109 & 2.71 & 2.18 & 0.504 & 184.15 & 8.36 & 20.85 \\ \hline
$C$ & 1.12 & 13.67 & 0.113 & 2.70 & 2.17 & 0.504 & 172.23 & 8.24 & 16.81 \\ \hline
$C$ & 1.12 & 13.67 & 0.119 & 2.68 & 2.16 & 0.504 & 145.60 & 6.24 & 14.24 \\ \hline
$C$ & 1.12 & 13.67 & 0.125 & 2.67 & 2.15 & 0.504 & 129.83 & 6.11 & 10.75 \\ \hline
$C$ & 1.12 & 13.67 & 0.133 & 2.65 & 2.13 & 0.504 & 99.01 & 4.60 & 7.17 \\ \hline
$C$ & 2.24 & 14.01 & 0.181 & 3.21 & 2.49 & 0.275 & 37.50 & 1.60 & 3.05 \\ \hline
$C$ & 2.24 & 14.01 & 0.219 & 3.16 & 2.45 & 0.275 & 44.01 & 1.85 & 3.75 \\ \hline
$C$ & 2.24 & 14.01 & 0.251 & 3.11 & 2.41 & 0.275 & 45.06 & 1.84 & 3.74 \\ \hline
$C$ & 2.20 & 13.17 & 0.292 & 2.17 & 1.77 & 0.642 & 27.97 & 1.35 & 3.47 \\ \hline
$C$ & 2.20 & 13.17 & 0.335 & 2.13 & 1.73 & 0.642 & 45.35 & 1.84 & 5.66 \\ \hline
$C$ & 2.20 & 13.17 & 0.363 & 2.09 & 1.71 & 0.642 & 55.11 & 2.75 & 6.97 \\ \hline
$C$ & 2.20 & 13.17 & 0.386 & 2.07 & 1.69 & 0.642 & 56.99 & 2.25 & 7.26 \\ \hline
$C$ & 4.73 & 15.03 & 0.397 & 4.59 & 3.27 & 0.263 & 2.42 & 0.16 & 0.30 \\ \hline
$C$ & 4.73 & 15.03 & 0.500 & 4.50 & 3.21 & 0.263 & 4.04 & 0.19 & 0.51 \\ \hline
$C$ & 3.94 & 14.77 & 0.320 & 4.23 & 3.08 & 0.391 & 5.49 & 0.22 & 0.35 \\ \hline
$C$ & 3.94 & 14.77 & 0.378 & 4.17 & 3.04 & 0.391 & 8.70 & 0.34 & 0.55 \\ \hline
$C$ & 3.94 & 14.77 & 0.429 & 4.12 & 3.00 & 0.391 & 10.04 & 0.34 & 0.62 \\ \hline
  \end{tabular}
  \end{center}
\end{small}
\caption{Extracted cross sections and their uncertainties for carbon
  target at eight kinematics 
  settings. The $Q^2$, W, $-t$, $P_{\pi}$ and $\epsilon$ values for
  each kinematics are shown at the same time. } 
\label{table:cdxs}
\end{table*}

\begin{table*}
  \begin{small}
  \begin{center}
    \begin{tabular}{|c|c|c|c|c|c|c|c|c|c|}
      \hline
      Target & $Q^2$ & W & -t & $P_{\pi}$  & $P^{CM}_{\pi}$ & $\epsilon$ &
      $\frac{d\sigma}{dtdP^{CM}_{\pi}}$ & stat. err. & sys. err.\\\hline
        & GeV$^2$ & GeV &  GeV$^2$ & GeV/c & GeV/c &  & $\mu b/ GeV^3$& 
      $\mu b/ GeV^3$ &
      $\mu b/ GeV^3$ \\\hline
$Cu$ & 2.24 & 62.36 & 0.161 & 3.23 & 3.05 & 0.558 & 81.49 & 3.06 & 6.15 \\ \hline
$Cu$ & 2.24 & 62.36 & 0.183 & 3.21 & 3.03 & 0.558 & 112.51 & 4.20 & 8.81 \\ \hline
$Cu$ & 2.24 & 62.36 & 0.197 & 3.19 & 3.01 & 0.558 & 118.28 & 4.48 & 9.30 \\ \hline
$Cu$ & 2.24 & 62.36 & 0.208 & 3.17 & 2.99 & 0.558 & 131.91 & 5.07 & 9.97 \\ \hline
$Cu$ & 2.24 & 62.36 & 0.219 & 3.16 & 2.98 & 0.558 & 131.81 & 4.85 & 9.93 \\ \hline
$Cu$ & 2.24 & 62.36 & 0.230 & 3.14 & 2.96 & 0.558 & 134.78 & 4.92 & 9.81 \\ \hline
$Cu$ & 2.24 & 62.36 & 0.241 & 3.12 & 2.95 & 0.558 & 125.55 & 4.81 & 9.23 \\ \hline
$Cu$ & 2.24 & 62.36 & 0.253 & 3.11 & 2.93 & 0.558 & 122.13 & 4.47 & 8.79 \\ \hline
$Cu$ & 2.24 & 62.36 & 0.265 & 3.09 & 2.91 & 0.558 & 112.68 & 4.25 & 8.12 \\ \hline
$Cu$ & 2.24 & 62.36 & 0.278 & 3.07 & 2.90 & 0.558 & 106.16 & 4.28 & 7.61 \\ \hline
$Cu$ & 4.04 & 63.12 & 0.364 & 3.98 & 3.71 & 0.257 & 10.98 & 0.86 & 1.51 \\ \hline
$Cu$ & 4.04 & 63.12 & 0.465 & 3.89 & 3.62 & 0.257 & 17.60 & 0.81 & 2.37 \\ \hline
$Cu$ & 3.02 & 62.64 & 0.250 & 3.51 & 3.29 & 0.450 & 32.07 & 1.37 & 3.98 \\ \hline
$Cu$ & 3.02 & 62.64 & 0.305 & 3.45 & 3.24 & 0.450 & 46.62 & 1.40 & 5.01 \\ \hline
$Cu$ & 3.02 & 62.64 & 0.363 & 3.38 & 3.17 & 0.450 & 47.83 & 1.61 & 4.74 \\ \hline
$Cu$ & 1.12 & 61.93 & 0.070 & 2.79 & 2.66 & 0.503 & 122.69 & 4.62 & 9.51 \\ \hline
$Cu$ & 1.12 & 61.93 & 0.083 & 2.76 & 2.63 & 0.503 & 618.86 & 18.04 & 49.79 \\ \hline
$Cu$ & 1.12 & 61.93 & 0.091 & 2.75 & 2.62 & 0.503 & 614.14 & 19.83 & 48.08 \\ \hline
$Cu$ & 1.12 & 61.93 & 0.098 & 2.73 & 2.60 & 0.503 & 552.29 & 18.40 & 43.14 \\ \hline
$Cu$ & 1.12 & 61.93 & 0.105 & 2.71 & 2.58 & 0.503 & 542.30 & 16.21 & 39.69 \\ \hline
$Cu$ & 1.12 & 61.93 & 0.115 & 2.69 & 2.56 & 0.503 & 484.09 & 14.21 & 33.64 \\ \hline
$Cu$ & 1.12 & 61.93 & 0.126 & 2.67 & 2.54 & 0.503 & 380.04 & 12.93 & 25.42 \\ \hline
$Cu$ & 2.24 & 62.35 & 0.186 & 3.19 & 3.01 & 0.282 & 109.18 & 4.45 & 9.03 \\ \hline
$Cu$ & 2.24 & 62.35 & 0.241 & 3.11 & 2.94 & 0.282 & 109.87 & 4.09 & 8.17 \\ \hline
$Cu$ & 4.73 & 63.70 & 0.468 & 4.53 & 4.18 & 0.264 & 10.76 & 0.38 & 1.40 \\ \hline
$Cu$ & 4.73 & 63.70 & 0.647 & 4.36 & 4.02 & 0.264 & 10.77 & 0.64 & 1.42 \\ \hline
$Cu$ & 3.94 & 63.34 & 0.314 & 4.24 & 3.93 & 0.391 & 14.16 & 0.62 & 1.77 \\ \hline
$Cu$ & 3.94 & 63.34 & 0.385 & 4.17 & 3.87 & 0.391 & 21.61 & 0.74 & 2.00 \\ \hline
$Cu$ & 3.94 & 63.34 & 0.450 & 4.10 & 3.80 & 0.391 & 23.31 & 0.75 & 1.94 \\ \hline
\end{tabular}
  \end{center}
\end{small}
\caption{Extracted cross sections and their uncertainties for copper
  target at seven kinematics 
  settings. The $Q^2$, W, $-t$, $P_{\pi}$ and $\epsilon$ values for
  each kinematics are shown at the same time. } 
\label{table:codxs}
\end{table*}

\begin{table*}
  \begin{small}
  \begin{center}
    \begin{tabular}{|c|c|c|c|c|c|c|c|c|c|}
      \hline
      Target & $Q^2$ & W & -t & $P_{\pi}$  & $P^{CM}_{\pi}$ & $\epsilon$ &
      $\frac{d\sigma}{dtdP^{CM}_{\pi}}$ & stat. err. & sys. err.\\\hline
        & GeV$^2$ & GeV &  GeV$^2$ & GeV/c & GeV/c &  & $\mu b/ GeV^3$& 
      $\mu b/ GeV^3$ &
      $\mu b/ GeV^3$ \\\hline
$Au$ & 2.16 & 186.68 & 0.164 & 3.23 & 3.17 & 0.557 & 201.07 & 7.93 & 24.37 \\ \hline
$Au$ & 2.16 & 186.68 & 0.221 & 3.15 & 3.09 & 0.557 & 247.23 & 6.02 & 29.46 \\ \hline
$Au$ & 4.03 & 187.48 & 0.438 & 3.92 & 3.82 & 0.256 & 33.28 & 3.04 & 3.15 \\ \hline
$Au$ & 3.02 & 186.97 & 0.254 & 3.50 & 3.43 & 0.449 & 63.50 & 3.79 & 9.00 \\ \hline
$Au$ & 3.02 & 186.97 & 0.323 & 3.43 & 3.36 & 0.449 & 90.51 & 3.56 & 12.80 \\ \hline
$Au$ & 1.12 & 186.22 & 0.070 & 2.79 & 2.74 & 0.504 & 149.26 & 6.50 & 14.36 \\ \hline
$Au$ & 1.12 & 186.22 & 0.083 & 2.76 & 2.72 & 0.504 & 1052.70 & 37.33 & 104.4 \\ \hline
$Au$ & 1.12 & 186.22 & 0.092 & 2.74 & 2.70 & 0.504 & 1085.12 & 37.20 &  109.3\\ \hline
$Au$ & 1.12 & 186.22 & 0.101 & 2.72 & 2.68 & 0.504 & 1033.13 & 36.08 & 103.5 \\ \hline
$Au$ & 1.12 & 186.22 & 0.110 & 2.70 & 2.66 & 0.504 & 908.06 & 34.59 & 85.81 \\ \hline
$Au$ & 1.12 & 186.22 & 0.118 & 2.68 & 2.64 & 0.504 & 881.15 & 34.71 & 75.30\\ \hline
$Au$ & 4.73 & 188.08 & 0.545 & 4.46 & 4.34 & 0.263 & 20.08 & 0.97 & 2.79 \\ \hline
$Au$ & 3.95 & 187.70 & 0.340 & 4.21 & 4.11 & 0.391 & 30.82 & 1.38 & 4.31 \\ \hline
$Au$ & 3.95 & 187.70 & 0.424 & 4.13 & 4.03 & 0.391 & 42.07 & 1.77 & 5.68 \\ \hline
    \end{tabular}
  \end{center}
\end{small}
\caption{Extracted cross sections and their uncertainties for gold
  target at six kinematics 
  settings. The $Q^2$, W, $-t$, $P_{\pi}$ and $\epsilon$ values for
  each kinematics are shown at the same time. } 
\label{table:gdxs}
\end{table*}

\begin{table*}
\begin{small}
  \begin{center}
    \begin{tabular}{|c|c|c|c|c|c|c|c|c|c|}
      \hline
      Target & $Q^2$ & W & -t & $P_{\pi}$  & $P^{CM}_{\pi}$ & $\epsilon$ &
      $\frac{d\sigma}{dtdP^{CM}_{\pi}}$ & stat. err. & sys. err.\\\hline
        & GeV$^2$ & GeV &  GeV$^2$ & GeV/c & GeV/c &  & $\mu b/ GeV^3$& 
      $\mu b/ GeV^3$ &
      $\mu b/ GeV^3$ \\\hline
$Al$ & 2.11 & 28.20 & 0.172 & 3.22 & 2.84 & 0.554 & 100.26 & 4.31 & 11.57 \\ \hline
$Al$ & 2.11 & 28.20 & 0.228 & 3.14 & 2.77 & 0.554 & 82.38 & 3.56 & 7.81 \\ \hline
$Al$ & 3.99 & 28.87 & 0.404 & 3.94 & 3.37 & 0.256 & 12.90 & 3.32 & 1.58 \\ \hline
$Al$ & 2.95 & 28.44 & 0.310 & 3.44 & 2.99 & 0.448 & 38.98 & 2.09 & 5.05 \\ \hline
$Al$ & 1.11 & 27.79 & 0.077 & 2.78 & 2.49 & 0.502 & 401.69 & 15.79 & 53.74 \\ \hline
$Al$ & 1.11 & 27.79 & 0.105 & 2.72 & 2.44 & 0.502 & 312.93 & 11.48 & 27.55 \\ \hline
$Al$ & 2.07 & 28.21 & 0.186 & 3.21 & 2.82 & 0.267 & 82.96 & 10.66 & 10.93 \\ \hline
$Al$ & 2.09 & 27.25 & 0.286 & 2.15 & 1.95 & 0.648 & 81.05 & 9.11 & 10.05 \\ \hline
$Al$ & 4.57 & 29.41 & 0.458 & 4.54 & 3.81 & 0.261 & 9.44 & 1.86 & 1.18 \\ \hline
$Al$ & 3.83 & 29.09 & 0.384 & 4.17 & 3.54 & 0.389 & 15.80 & 1.36 & 2.04 \\ \hline
  \end{tabular}
  \end{center}
\end{small}
\caption{Extracted cross sections and their uncertainties for aluminum
  target at eight kinematics 
  settings. The $Q^2$, W, $-t$, $P_{\pi}$ and $\epsilon$ values for
  each kinematics are shown at the same time. } 
\label{table:aldxs}
{\bf{\large Appendix C}}
  \begin{small}
  \begin{center}
    \begin{tabular}{|c|c|c|c|c|c|c|c|c|c|c|c|c|c|}
      \hline
      Target & $Q^2$ &  $P_{\pi}$  & $k_{\pi}$ & $\epsilon$ &
      T& stat.   & sys.   & $T_{A,2}$ & stat.   & sys.  & $T_{A,12}$ & stat.   & sys.   \\\hline
        & GeV$^2$ & GeV/c &  GeV/c &  & & & & & & & & &\\\hline
$LD_2$ & 1.1 & 2.8 & 0.23 & 0.50 & 0.98 & 0.02 & 0.03 & - & - & - & - & - & - \\ \hline
$LD_2$ &2.2 & 3.2 & 0.41 & 0.56 & 1.01 & 0.02 & 0.03 & - & - & - & - & - & - \\ \hline
$LD_2$ &3.0 & 3.4 & 0.56 & 0.45 & 0.99 & 0.02 & 0.04 & - & - & - & - &- & - \\ \hline
$LD_2$ &3.9 & 4.1 & 0.70 & 0.39 & 1.05 & 0.02 & 0.04 & - & - & - & - & - & - \\ \hline
$LD_2$ &4.7 & 4.4 & 0.79 & 0.26 & 1.03 & 0.03 & 0.04 & - & - & - & - & - & - \\ \hline
$LD_2$ &2.2 & 3.2 & 0.42 & 0.27 & 1.04 & 0.03 & 0.03 & - & - & - & - & - & - \\ \hline
$LD_2$ &4.0 & 3.9 & 0.71 & 0.25 & 1.07 & 0.04 & 0.04 & - & - & - & - & - & - \\ \hline
$LD_2$ &2.2 & 2.1 & 0.65 & 0.63 & 1.00 & 0.02 & 0.03 & - & - & - & - & - & - \\ \hline
$C$ &1.1 & 2.8 & 0.23 & 0.50 & 0.67 & 0.01 & 0.02 & 0.68 & 0.01 & 0.02 & - & -& - \\ \hline
$C$ &2.2 & 3.2 & 0.41 & 0.56 & 0.65 & 0.01 & 0.02 & 0.65 & 0.01 & 0.02 & - & - & - \\ \hline
$C$ &3.0 & 3.4 & 0.56 & 0.45 & 0.68 & 0.02 & 0.03 & 0.68 & 0.02 & 0.03 & - & - & - \\ \hline
$C$ &3.9 & 4.1 & 0.70 & 0.39 & 0.77 & 0.02 & 0.03 & 0.73 & 0.02 & 0.03 & - & - & - \\ \hline
$C$ &4.7 & 4.4 & 0.79 & 0.26 & 0.70 & 0.03 & 0.03 & 0.69 & 0.03 & 0.03 & - & - & - \\ \hline
$C$ &2.2 & 3.2 & 0.42 & 0.27 & 0.60 & 0.02 & 0.02 & 0.57 & 0.02 & 0.02 & - & - & - \\ \hline
$C$ &4.0 & 3.9 & 0.71 & 0.25 & 0.68 & 0.03 & 0.03 & 0.64 & 0.03 & 0.03 & - & - & - \\ \hline
$C$ &2.2 & 2.1 & 0.65 & 0.63 & 0.66 & 0.02 & 0.02 & 0.66 & 0.02 & 0.02 & - & - & - \\ \hline
$Al$ &1.1 & 2.8 & 0.23 & 0.50 & 0.49 & 0.01 & 0.02 & 0.50 & 0.02 & 0.02 & 0.73 & 0.02 & 0.03 \\ \hline
$Al$ &2.2 & 3.2 & 0.41 & 0.56 & 0.52 & 0.02 & 0.02 & 0.52 & 0.02 & 0.02 & 0.80 & 0.03 & 0.03 \\ \hline
$Al$ &3.0 & 3.4 & 0.56 & 0.45 & 0.57 & 0.03 & 0.02 & 0.57 & 0.03 & 0.02 & 0.83 & 0.05 & 0.03 \\ \hline
$Al$ &3.9 & 4.1 & 0.70 & 0.39 & 0.59 & 0.05 & 0.03 & 0.56 & 0.05 & 0.03 & 0.77 & 0.07 & 0.03 \\ \hline
$Al$ &4.7 & 4.4 & 0.79 & 0.26 & 0.71 & 0.14 & 0.03 & 0.69 & 0.14 & 0.03 & 1.01 & 0.20 & 0.04 \\ \hline
$Al$ &2.2 & 3.2 & 0.42 & 0.27 & 0.46 & 0.06 & 0.02 & 0.45 & 0.06 & 0.02 & 0.78 & 0.10 & 0.03 \\ \hline
$Al$ &4.0 & 3.9 & 0.71 & 0.25 & 0.70 & 0.16 & 0.03 & 0.66 & 0.15 & 0.02 & 1.03 & 0.24 & 0.03 \\ \hline
$Al$ &2.2 & 2.1 & 0.65 & 0.63 & 0.59 & 0.07 & 0.02 & 0.60 & 0.07 & 0.02 & 0.90 & 0.10 & 0.03 \\ \hline
$Cu$ &1.1 & 2.8 & 0.23 & 0.50 & 0.45 & 0.01 & 0.01 & 0.46 & 0.01 & 0.02 & 0.67 & 0.01 & 0.02 \\ \hline
$Cu$ &2.2 & 3.2 & 0.41 & 0.56 & 0.45 & 0.01 & 0.02 & 0.45 & 0.01 & 0.02 & 0.70 & 0.01 & 0.03 \\ \hline
$Cu$ &3.0 & 3.4 & 0.56 & 0.45 & 0.43 & 0.01 & 0.02 & 0.44 & 0.01 & 0.02 & 0.64 & 0.02 & 0.03 \\ \hline
$Cu$ &3.9 & 4.1 & 0.70 & 0.39 & 0.52 & 0.01 & 0.02 & 0.49 & 0.01 & 0.02 & 0.67 & 0.02 & 0.03 \\ \hline
$Cu$ &4.7 & 4.4 & 0.79 & 0.26 & 0.53 & 0.02 & 0.02 & 0.51 & 0.02 & 0.02 & 0.75 & 0.04 & 0.03 \\ \hline
$Cu$ &2.2 & 3.2 & 0.42 & 0.27 & 0.43 & 0.02 & 0.02 & 0.42 & 0.01 & 0.02 & 0.73 & 0.03 & 0.03 \\ \hline
$Cu$ &4.0 & 3.9 & 0.71 & 0.25 & 0.51 & 0.02 & 0.02 & 0.48 & 0.02 & 0.02 & 0.75 & 0.04 & 0.03 \\ \hline
$Au$ &1.1 & 2.8 & 0.23 & 0.50 & 0.28 & 0.01 & 0.01 & 0.28 & 0.01 & 0.01 & 0.41 & 0.01 & 0.02 \\ \hline
$Au$ &2.2 & 3.2 & 0.41 & 0.56 & 0.29 & 0.01 & 0.01 & 0.28 & 0.01 & 0.01 & 0.44 & 0.01 & 0.02 \\ \hline
$Au$ &3.0 & 3.4 & 0.56 & 0.45 & 0.29 & 0.01 & 0.01 & 0.30 & 0.01 & 0.01 & 0.43 & 0.02 & 0.02 \\ \hline
$Au$ &3.9 & 4.1 & 0.70 & 0.39 & 0.34 & 0.01 & 0.02 & 0.32 & 0.01 & 0.01 & 0.44 & 0.02 & 0.02 \\ \hline
$Au$ &4.7 & 4.4 & 0.79 & 0.26 & 0.33 & 0.02 & 0.02 & 0.32 & 0.02 &
      0.01 & 0.46 & 0.03 & 0.02 \\ \hline
$Au$ &4.0 & 3.9 & 0.71 & 0.25 & 0.31 & 0.03 & 0.02 & 0.29 & 0.03 & 0.01 & 0.46 & 0.05 & 0.02 \\ \hline
    \end{tabular}
  \end{center}
\end{small}
\caption{Extracted nuclear transparency and their uncertainties.
  The $Q^2$, $P_{\pi}$, $k_{\pi}$ and $\epsilon$ values for
  each kinematics are shown at the same time. Here the $T$, $T_{A,2}$
  and $T_{A,12}$ are the nuclear transparency formed with hydrogen,
  deuterium and carbon targets, respectively. The $Q^2$ dependent
  model uncertainty is 7.6\%, 5.7\%, 3.5\%, 3.8\%, and 3.8\% for $Q^2$
  = 1.1, 2.1, 3.0, 3.9, 4.7 (GeV/c)$^2$, respectively.} 
\label{table:tlh2}
\end{table*}

\end{document}